\documentclass[twoside]{IEEEtran}
\usepackage{ifpdf}
\usepackage{cite}
\ifCLASSINFOpdf
  \usepackage[pdftex]{graphicx}
  \graphicspath{{figs/}{}}
  \DeclareGraphicsExtensions{.pdf,.jpeg,.png}
  \usepackage{epstopdf}
\else
  \usepackage[dvips]{graphicx}
  \graphicspath{{figs/}}
  \DeclareGraphicsExtensions{.eps}
\fi
\usepackage[cmex10]{amsmath}
\usepackage{amssymb}

\usepackage{array}

\usepackage[normalem]{ulem}	% use for strikethrough text
\usepackage{verbatim}

\usepackage[dvipsnames]{xcolor}
\usepackage{version}
\usepackage{siunitx}
\usepackage{bm}
\usepackage{xspace}
\usepackage[normalem]{ulem}

\usepackage{tikz}
\usepackage{textcomp}
\usepackage{hyperref}

\excludeversion{notes}

\newcommand\copyrighttext{%
  \footnotesize \textcopyright 2021 IEEE. Personal use of this material is permitted.
  Permission from IEEE must be obtained for all other uses, in any current or future
  media, including reprinting/republishing this material for advertising or promotional
  purposes, creating new collective works, for resale or redistribution to servers or
  lists, or reuse of any copyrighted component of this work in other works.
  }
\newcommand\copyrightnotice{%
\begin{tikzpicture}[remember picture,overlay]
\node[anchor=south,yshift=3pt] at (current page.south) {\fbox{\parbox{\dimexpr\textwidth-\fboxsep-\fboxrule\relax}{\copyrighttext}}};
\end{tikzpicture}%
}

\begin{document}
\title{Mid-Range Wireless Power Transfer at 100~MHz using Magnetically-Coupled Loop-Gap Resonators}

%\thanks{M. Shell is with the Department of Electrical and Computer Engineering, Georgia Institute of Technology, Atlanta, GA, 30332 USA e-mail: (see http://www.michaelshell.org/contact.html).}% 

\author{David~M.~Roberts, Aaron~P.~Clements, Rowan~McDonald, Jake~S.~Bobowski, and Thomas~Johnson,~\IEEEmembership{Member,~IEEE}
       %~\IEEEmembership{Member,~IEEE}%,~\IEEEmembership{Life~Fellow,~IEEE}% <-this % stops a space
\thanks{Manuscript submitted on \today.}
\thanks{D.M.~Roberts and J.S.~Bobowski are with the Department
of Physics, University of British Columbia, Kelowna,
BC, V1V 1V7 Canada (e-mail: \mbox{jake.bobowski@ubc.ca}).}% <-this % stops a space
\thanks{A.P.~Clements, R.~McDonald, and T.~Johnson are with the School of Engineering, University of British Columbia, Kelowna, BC, V1V 1V7, Canada (e-mail: \mbox{thomas.johnson@ubc.ca}).}% <-this % stops a space
}



%\markboth{IEEE Transactions on Microwave Theory and Techniques}%
%{Roberts \MakeLowercase{\textit{et al.}}: Mid-Range Wireless Power Transfer at 100~MHz using Magnetically-Coupled Loop-Gap Resonators}
\markboth{}{Roberts \MakeLowercase{\textit{et al.}}: Mid-Range Wireless Power Transfer at 100~MHz using Magnetically-Coupled Loop-Gap Resonators}



\maketitle
\copyrightnotice

\begin{abstract}
We describe efficient four-coil inductive power transfer (IPT) systems that operate at 100~MHz.  The magnetically-coupled transmitter and receiver were made from electrically-small and high-$\boldsymbol{Q}$ loop-gap resonators (LGRs).  In contrast to the commonly-used helical and spiral resonators, the LGR design has the distinct advantage that electric fields are strongly confined to the capacitive gap of the resonator.  With negligible fringing electric fields in the surrounding space, the IPT system is immune to interference from nearby dielectric objects, even when they are in close proximity to the transmitter and/or receiver.  We experimented with both cylindrical and split-toroidal LGR geometries.  Although both systems performed well under laboratory conditions, the toroidal geometry has the additional advantage that the magnetic flux is weak everywhere except within the bore of the LGR and in the space directly between the transmitter and receiver.  Furthermore, we show that the toroidal LGR system can be operated efficiently at a fixed frequency for a wide range of transmitter-receiver distances.  The experimental results are complimented by 3-D finite-element simulations which were used to investigate the electromagnetic field profiles and surface current density distributions.  Finally, we demonstrate the use of our IPT system at powers up to 32~W and discuss possible applications.
\end{abstract}


% \fi
\IEEEpeerreviewmaketitle

\section{\label{sec:intro}Introduction}

\IEEEPARstart{M}{ost} inductive power transfer (IPT) systems use helical~\cite{Soljacic:2007, RamRakhyani:2011, Lee:2011, Cheon:2011, Li:2012, Delichte:2018} or spiral resonators~\cite{Sample:2011, Park:2011, Duong:2011, Sample:2013, Duong:2015}.  These types of resonators are easy to construct, have long free-space wavelengths $\lambda$ compared to the resonator diameter $d$ with $d/\lambda\sim 0.02$, and have high quality factors $Q\sim 10^3$~\cite{Soljacic:2007}.    One drawback, however, is that the electric fields produced in the vicinity of these resonators can be substantial.  As a result, the power transfer efficiency is susceptible to degradation due to nearby non-resonant dielectric objects.  The magnetic fields, while most intense between the transmitter and receiver, also occupy the complete volume of space surrounding the system~\cite{Dioniqi:2015, Park:2013, Park:2016}.  On the one hand, this magnetic field configuration allows for mid-range IPT that is to some extent omnidirectional~\cite{Karalis:2008, Sample:2013}.  On the other hand, IPT systems using helical and spiral resonators produce near-field electromagnetic (EM) waves of strengths that, even for modest powers, quickly approach the recommended human exposure limits~\cite{IEEEstd:2006, ICNIRP:2020}.

Lorenz and coworkers have designed intricate coil geometries for spiral-type resonators that suppress losses associated with both the skin effect in conductors and the proximity effect in which the magnetic fields of adjacent windings interact with one another~\cite{S-HLee:2011, Zhu:2019}.  This same group has also used their novel coil designs to reduce the electric and magnetic field strengths in the regions between the transmitter and receiver coils while simultaneously maintaining high power transfer efficiency~\cite{Zhu:2018a}.  Furthermore, in an effort to develop practical high-power IPT systems that comply with safety standards, they have proposed ferrite shield geometries that could be used to further reduce EM field strengths in the space between the transmitter and receiver~\cite{Zhu:2018b, Zhu:2020}.

There are also examples of IPT systems that have been implemented using alternative resonator designs.  Perhaps most notably, Chabalko {\it et al.}\ have demonstrated a system in which an entire room has been fashioned into a cavity resonator that acts as the transmitter.  Power can be transferred with good efficiency to multiple devices distributed throughout the room~\cite{Chabalko:2014, Chabalko:2017}.  Dionigi and Mongiardo have constructed a loop resonator using a length of coaxial cable~\cite{Dionigi:2012}.  The electric fields are predominantly confined to the interior of the coaxial cable while the external magnetic field is due to surface currents on the outer conductor.  Song {\it et al.}\ have coupled the magnetic dipoles induced in low-loss dielectric resonators of Mg-doped $\left(\mathrm{Ba},\mathrm{Sr}\right)\mathrm{TiO}_3$ with relative permittivity $\varepsilon_\mathrm{r}\approx 10^3$~\cite{Song:2016}.  Finally, metamaterials designed to have a negative effective permeability have been used in an effort to enhance the power transfer efficiency and extend the range of magnetically-coupled wireless power transfer (WPT) systems~\cite{Wang:2011, Urzhumov:2011, Lipworth:2014, Wang:2013}.

In this paper, our objective was to develop an IPT system in which the electric fields are completely confined within the resonators and the external magnetic field is weak everywhere except in the space between transmitter and receiver.  Such a system would be insensitive to the presence of nearby dielectric materials, be insensitive to nearby conductors provided that they were not directly between the two resonators, and could potentially be used to limit human exposure to RF EM fields in high-power applications.

To realize these objectives, we use magnetically-coupled loop-gap resonators (LGRs) to achieve efficient mid-range IPT for frequencies in the range of \SI{100}{\mega\hertz}.  The design of LGRs is well documented~\cite{Hardy:1981, Froncisz:1982}. LGRs of various geometries have been used as electron spin resonance (ESR) spectrometers~\cite{Wood:1984, Froncisz:1986, Hyde:2007} and to measure the electric~\cite{Bobowski:2013, Bobowski:2017} and magnetic~\cite{Bonn:1991, Hardy:1993, Dubreuil:2019, Bobowski:2018, Madsen:2020} properties of a wide variety of materials.  The resonator consists of a physical structure that has a cylindrical cavity  split by a gap that runs the length of the cylinder. Currents on the inside surface of the cylinder generate an inductive reactance, while the electric field across the gap generates a capacitive reactance. The physical geometry of the cylinder and the gap can be designed to implement  high-$Q$ resonators. One of the advantages of LGRs is that the narrow capacitive gap desensitizes the resonator to nearby dielectric materials. The cylindrical LGR (CLGR) can also be configured as a torus to implement a toroidal LGR (TLGR) that has the additional advantage of confining the magnetic field to the interior cavity with negligible leakage to the outside~\cite{Bobowski:2016}. The magnetic field confinement of the TLGR provides additional desensitization to nearby magnetic materials that can introduce power loss in IPT systems.

The contributions of this work were: (1) to show experimental results for an innovative IPT system that uses LGRs to efficiently couple power through a medium, (2) to show EM simulation results that demonstrate the shaping and confinement of magnetic and electric fields in the region surrounding the LGRs in an IPT configuration and to show that field confinement is significantly better using TLGRs compared to CLGRs, (3) to show experimental results that demonstrate a wide fixed-frequency spatial bandwidth that is unique to the TLGR geometry, and (4) to describe an experimental method that can be used to independently determine all three of the dominant coupling coefficients in a four-coil IPT system. We also expect that our work will stimulate new research to evaluate the potential advantages of using LGR transmitters and receivers in high-power IPT applications.

The paper is organized as follows: Section~\ref{sec:methods} introduces equivalent circuits and data analysis methods used to interpret the experimental results.  Section~\ref{sec:LGRs} describes LGRs and emphasizes the unique characteristics that make them a good fit for IPT applications.  In Section~\ref{sec:sims}, detailed numerical simulations are used to investigate the EM field configurations and surface current densities of the two LGR IPT designs.  Section~\ref{sec:expt} presents the experimental measurements and data analysis.  The implications of the experimental and numerical results are discussed within the context of potential applications in Section~\ref{sec:discussion}.  Finally, Section~\ref{sec:conclusions} summarizes the main conclusions.

\section{\label{sec:methods}Data Analysis Methods}
This section describes the methods used to analyze the scattering parameter data obtained from our IPT systems.  In particular, equivalent circuits are used to develop models that can be fit to reflection coefficient ($S_{11}$) and transmission coefficient ($S_{21}$) measurements.  These fits are used to extract the primary coupling coefficients relevant to four-coil IPT systems.  Although this paper focuses on IPT using LGR transmitters and receivers, the analysis methods presented in this section can be applied to any four-coil IPT system. 

\subsection{\label{sec:circuit}Equivalent-Circuit Model}
\begin{figure}
\begin{tabular}{c}
(a)~\includegraphics[width=0.8\columnwidth]{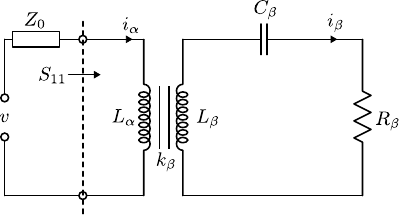}\\
~\\
(b)~\includegraphics[width=0.92\columnwidth]{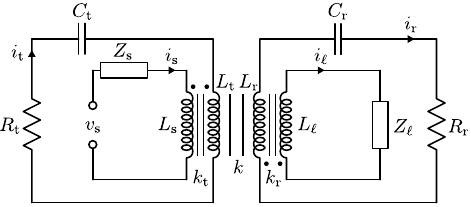}
\end{tabular}
\caption{\label{fig:circuits}Equivalent circuit models used to analyze measured scattering parameters. (a) A series resonator ($L_{\beta}, C_{\beta}, R_{\beta})$ coupled to an inductive loop $L_{\alpha}$ that is driven by a signal generator $v$ with output impedance $Z_0$.  The dashed line represents a VNA calibration plane.  The equivalent circuit models the transmit or receive resonator in an IPT system. (b) Equivalent circuit of a full IPT system.  The model assumes that the source loop only couples to the transmit resonator (coupling coefficient $k_\mathrm{t}$) and the load loop only couples to the receive resonator (coupling coefficient $k_\mathrm{r}$). The separation distance $x$ between the transmitter and receiver is modeled by coupling coefficient $k$.}
\end{figure}

Figure~\ref{fig:circuits}(a) is an equivalent circuit model of a loop of inductance $L_\alpha$ magnetically coupled to a series $LRC$ resonator.  The coupling loop is driven by a signal generator with output impedance $Z_0$.  This circuit represents either a source ($\alpha = \mathrm{s}$) coupling loop coupled to a resonant transmitter ($\beta =\mathrm{t}$) or a load ($\alpha = \ell$) coupling loop coupled to a resonant receiver ($\beta =\mathrm{r}$).  In this figure, it is assumed that the transmitter and receiver have been isolated from one another.  The relevant coupling coefficient $k_\beta$ is determined from the mutual inductance \mbox{$M_\beta=k_\beta\sqrt{L_\alpha L_\beta}$} between $L_\alpha$ and $L_\beta$.  

The circuit in Fig.~\ref{fig:circuits}(a) was first analyzed in \cite{Rinard:1993} and subsequently used to calculate the reflection coefficient $S_{11}$ at the signal generator reference plane~\cite{Bobowski:2018}.  $S_{11}$ can be expressed in terms of the effective impedance $Z$ of the coupled resonator
\begin{equation}
S_{11}=\frac{Z-Z_0}{Z+Z_0},\label{eq:S11}
\end{equation}
where the real and imaginary components of $Z=R_Z+jX_Z$ are given by:
\begin{align}
R_Z &=\frac{k_\beta^2\omega L_\alpha Q_\beta\dfrac{\omega}{\omega_\beta}\sqrt{\dfrac{\omega}{\omega_\beta}}}{\dfrac{\omega}{\omega_\beta}+Q_\beta^2\left(\dfrac{\omega}{\omega_\beta}-\dfrac{\omega_\beta}{\omega}\right)^2}\label{eq:R}\\
X_Z &=\omega L_\alpha\left[1-\frac{k_\beta^2 Q_\beta^2\dfrac{\omega}{\omega_\beta}\left(\dfrac{\omega}{\omega_\beta}-\dfrac{\omega_\beta}{\omega}\right)}{\dfrac{\omega}{\omega_\beta}+Q_\beta^2\left(\dfrac{\omega}{\omega_\beta}-\dfrac{\omega_\beta}{\omega}\right)^2}\right].\label{eq:X}
\end{align}
In these expressions, \mbox{$\omega_\beta^{-2}=L_\beta C_\beta$} and \mbox{$Q_\beta^{-1}=R_\beta\sqrt{C_\beta/L_\beta}$} represent the unloaded (i.e.\ the $k_\beta\to 0$ limit) resonant frequency and quality factor of the resonator, respectively.  The $\sqrt{\omega/\omega_\beta}$ factor in (\ref{eq:R}) is due to the frequency dependence of the effective resistance of the resonator which is determined by the skin depth $\delta=\sqrt{2\rho/\left(\mu_0\omega\right)}$, where $\rho$ is resistivity, $\omega$ is angular frequency, and $\mu_0$ is the permeability of free space~\cite{Bobowski:2013}.  

Equations (\ref{eq:S11}) - (\ref{eq:X}) can be used to determine $\left\vert S_{11}\right\vert$ for a resonator magnetically coupled to a loop of inductance $L_\alpha$.  As described in \cite{Bobowski:2018}, $L_\alpha$ can be determined from a separate measurement of $S_{11}$ taken from the coupling loop after it has been isolated from the resonator. With $L_\alpha$ known, fits to the measured $\left\vert S_{11}\right\vert$ spectra can be used to determine $\omega_\beta$, $Q_\beta$, and $k_\beta$. 

Figure~\ref{fig:circuits}(b) shows an equivalent circuit model of the four-coil IPT system.  In the model, cross-couplings are assumed to be negligible such that the source loop couples only to the transmitting resonator ($k_\mathrm{t}$) and the load loop couples only to the receiving resonator ($k_\mathrm{r}$).  The coupling between the transmitter and receiver is denoted $k$.  Analyses of similar equivalent circuits have been reported previously~\cite{Sample:2011, Cheon:2011, Duong:2011}.  In those cases, a discrete capacitance is placed in series with the coupling loops to make all four circuits resonant.  We have found that the additional capacitance is not needed to tune our IPT systems.  To avoid introducing additional losses, series capacitors are not used in our source and load loops.  

A Kirchhoff loop analysis of the circuits in Fig.~\ref{fig:circuits}(b) results in the following system of equations that can be solved for the unknown current phasors $I_\mathrm{s}$, $I_\mathrm{t}$, $I_\mathrm{r}$, and $I_\ell$
\begin{align}
0&=j\omega k_\mathrm{t} \sqrt{L_\mathrm{s} L_\mathrm{t}} I_\mathrm{s}-j\omega k \sqrt{L_\mathrm{t}L_\mathrm{r}} I_\mathrm{r}- I_\mathrm{t}Z_\mathrm{t},\label{eq:sysA}\\
0&=j\omega k_\mathrm{r}\sqrt{L_\ell L_\mathrm{r}} I_\ell-j\omega k \sqrt{L_\mathrm{t}L_\mathrm{r}} I_\mathrm{t}- I_\mathrm{r}Z_\mathrm{r},\label{eq:sysB}\\
0&= V_\mathrm{s}+j\omega k_\mathrm{t}\sqrt{L_\mathrm{s}L_\mathrm{t}}I_\mathrm{t}-\left(Z_0+j\omega L_{\mathrm{s}}\right) I_\mathrm{s},\label{eq:sys1}\\
0&=j\omega k_\mathrm{r}\sqrt{L_\ell L_\mathrm{r}}I_\mathrm{r}-\left(Z_0+j\omega L_{\ell}\right) I_{\ell},\label{eq:sys2}
\end{align}
where it has been assumed that $Z_\mathrm{s}=Z_\ell\equiv Z_0$ and the quantity $Z_\beta\equiv R_\beta+j\omega L_\beta+1/\left(j\omega C_\beta\right)$ has been defined.  

For the purposes of fitting measured $\left\vert S_{21}\right\vert$ spectra, it is convenient to re-express (\ref{eq:sysA}) and (\ref{eq:sysB}) in terms of of $\omega_\beta$ and $Q_\beta$ such that
\begin{align}
0&=jk_\mathrm{t}\frac{\omega}{\omega_{\mathrm{t}}} \sqrt{\frac{L_{\mathrm{s}}}{L_\mathrm{t}}} I_\mathrm{s}-jk \frac{\omega}{\sqrt{\omega_{\mathrm{t}}\omega_{\mathrm{r}}}}\sqrt{\frac{Q_{\mathrm{r}}}{Q_{\mathrm{t}}}} I_\mathrm{r}-\left(p_\mathrm{t}+jq_\mathrm{t}\right) I_\mathrm{t},\label{eq:sys3}\\
0&=jk_\mathrm{r}\frac{\omega}{\omega_{\mathrm{r}}} \sqrt{\frac{L_\ell}{L_\mathrm{r}}} I_\ell-jk \frac{\omega}{\sqrt{\omega_{\mathrm{t}}\omega_{\mathrm{r}}}}\sqrt{\frac{Q_{\mathrm{t}}}{Q_{\mathrm{r}}}} I_\mathrm{t}-\left(p_\mathrm{r}+jq_\mathrm{r}\right) I_\mathrm{r},\label{eq:sys4}
\end{align}
where the quantities \mbox{$p_\beta\equiv Q_\beta^{-1}\sqrt{\omega/\omega_\beta}$} and \mbox{$q_\beta\equiv\left(\omega/\omega_\beta-\omega_\beta/\omega\right)$} have been defined.  Solving (\ref{eq:sys1}) -- (\ref{eq:sys4}) for the current in the load loop $I_\ell$ allows one to calculate the scattering parameter \mbox{$S_{21}=2I_\ell Z_0/V_\mathrm{s}$}, which can be fit to experimental data to determine the coupling coefficient $k$ between the transmitter and receiver. 

\subsection{\label{sub:ExptMethod}Experimental Methods}
In Section~\ref{sub:exptCopmpare}, the efficiencies of both the CLGR- and  TLGR-based IPT systems are characterized as a function of the distance $x$ between transmitter and receiver.  For each value of $x$, the couplings $k_\mathrm{t}$ and $k_\mathrm{r}$ were tuned to achieve the optimal power transfer and the $\left\vert S_{21}\right\vert$ versus frequency data were recorded using a vector network analyzer (VNA). All scattering parameters reported in this paper were measured using an SDR-Kits DG8SAQ VNA.

Next, the response of the equivalent circuit described in Section~\ref{sec:circuit} is compared to the response of the TLGR IPT system. The equivalent circuit parameter values were obtained from a combination of analytic and experimental results, and the method is now summarized. For each value of $x$, after recording the optimal $\left\vert S_{21}\right\vert$ curve, the transmitting and receiving resonators were isolated from one another without changing the positions of the source and load coupling loops.  The VNA was then used to measure the $\left\vert S_{11}\right\vert$ frequency response of both the transmitter and receiver.  Equations (\ref{eq:S11}) -- (\ref{eq:X}), based on the equivalent circuit of Fig.~\ref{fig:circuits}(a), were then used to determine $\omega_\mathrm{t}$, $Q_\mathrm{t}$, $k_\mathrm{t}$, $\omega_\mathrm{r}$, $Q_\mathrm{r}$, and $k_\mathrm{r}$.  Finally, the $\left\vert S_{21}\right\vert$ data were fit to $2\left\vert I_\ell/V_\mathrm{s}\right\vert Z_0$ with $I_\ell$ determined from (\ref{eq:sys1}) -- (\ref{eq:sys4}) based on the equivalent circuit model of Fig.~\ref{fig:circuits}(b).  

This analysis allows for a determination of the $x$-dependencies of $k$, $k_\mathrm{t}$, and $k_\mathrm{r}$ for a tuned IPT system.  It also enables an experimental investigation of the relationship between the various coupling coefficients.  The coupling coefficient analysis for the TLGR IPT system is presented in Section~\ref{sec:expt}.  A similar analysis was not done for the CLGR system because our testbed did not allow the transmitting and receiving resonators to be easily isolated from one another while maintaining constant values of $k_\mathrm{t}$ and $k_\mathrm{r}$.

\section{\label{sec:LGRs}Loop-Gap Resonators}
\begin{figure*}
\begin{tabular}{cc}
(a)~\includegraphics[height = 4.22 cm]{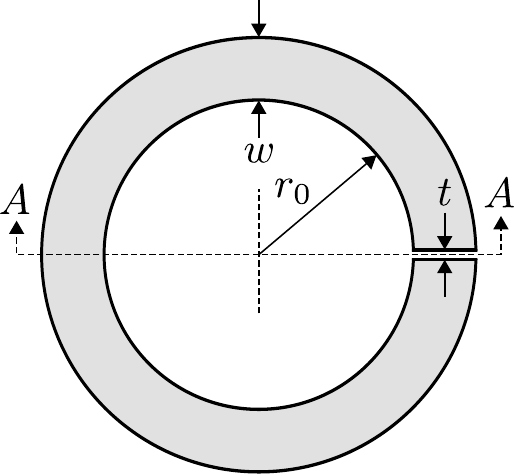} &
(b)~\includegraphics[height = 4.22 cm]{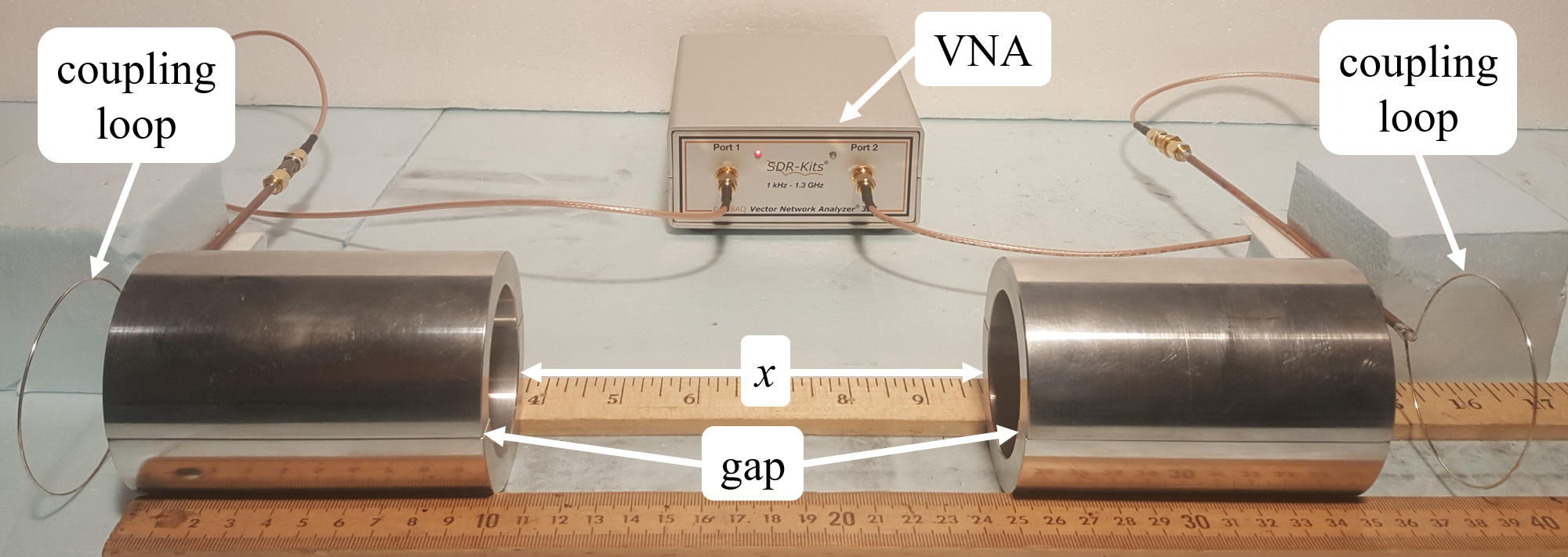}\\
~ & ~\\
\end{tabular}
\begin{tabular}{cc}
(c)~\includegraphics[height = 4.65cm]{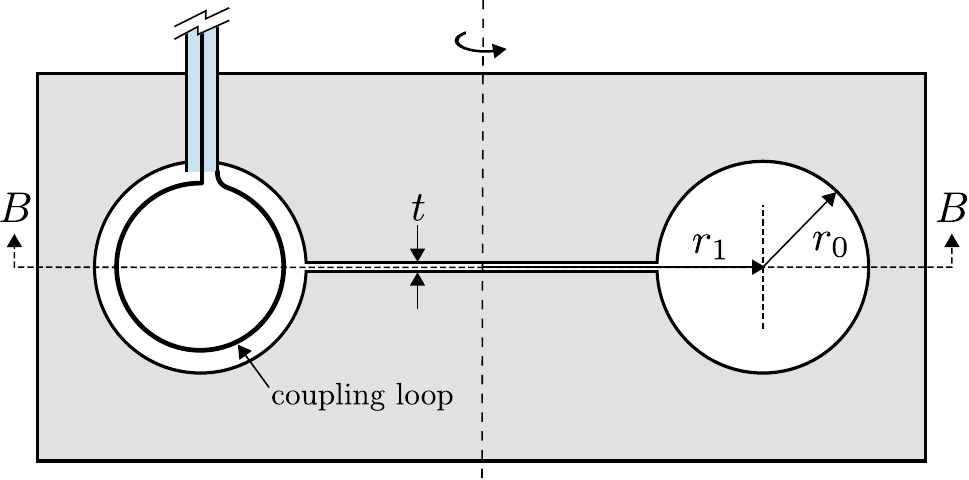} &
(d)~\includegraphics[height = 4.65cm]{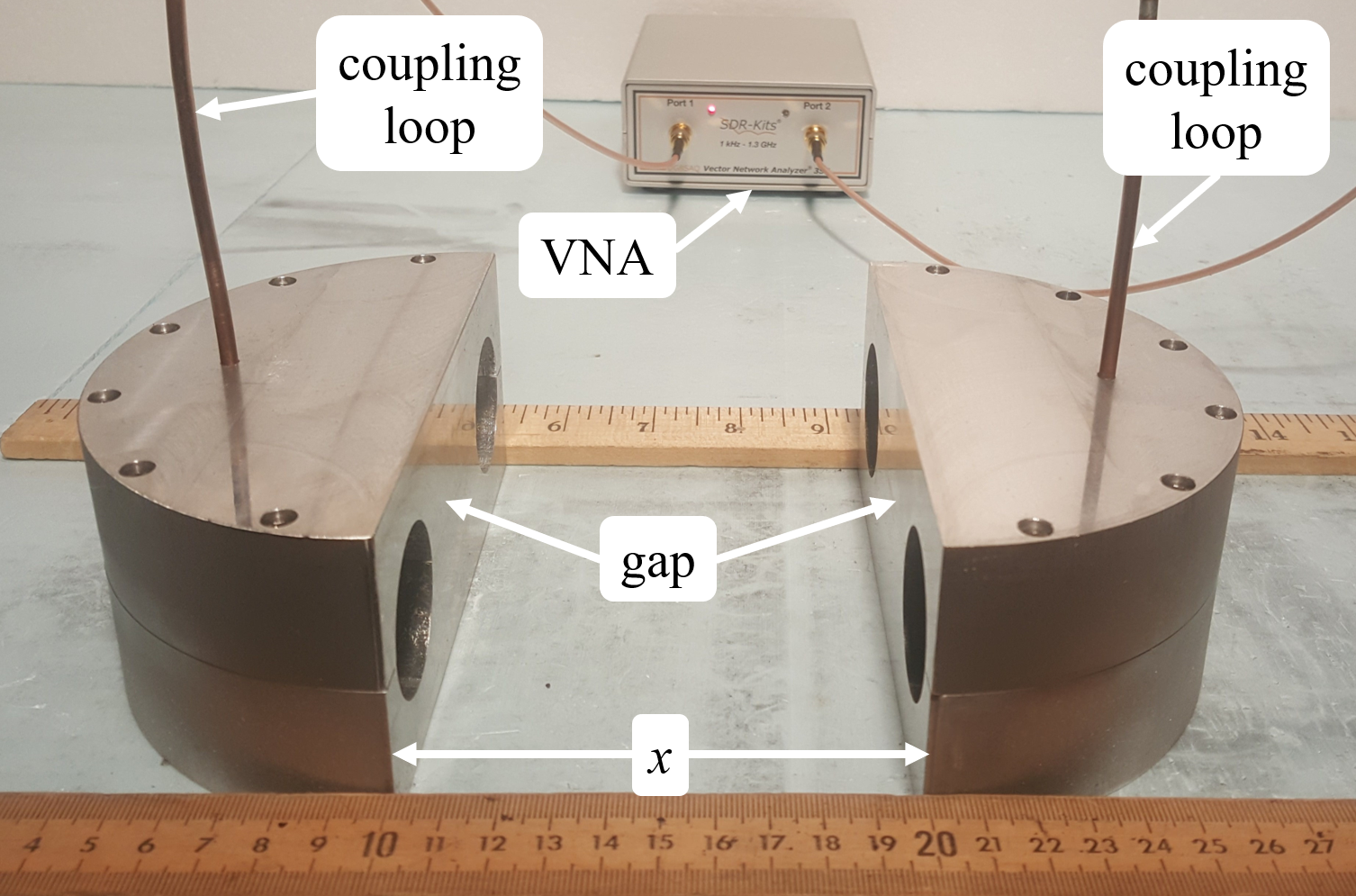}
\end{tabular}
\caption{\label{fig:LGRs}(a) Scale drawing of the CLGR cross-section with the gap size $t$ exaggerated by a factor of $10^3$.  The dimensions of the CLGR are $r_0=\SI{2.54}{\centi\meter}$, $w=\SI{1.02}{\centi\meter}$, $t=\SI{102}{\micro\meter}$, and its length is $\ell=\SI{10.16}{\centi\meter}$. (b) Photograph of the IPT experimental setup using identical CLGRs made from aluminum.  The source and load coupling loops, made from semi-rigid coaxial cable and attached to the two ports of the VNA, are positioned to achieve optimal power transfer efficiency. For scale, the transmitter and receiver are separated by a distance $x=\SI{15}{\centi\meter}$. (c) Scale drawing of the TLGR cross-section with the gap size $t$ exaggerated by a factor of $10^3$.  The dimensions of the TLGR are $r_0=\SI{1.75}{\centi\meter}$, $r_1=\SI{4.60}{\centi\meter}$, and $t=\SI{102}{\micro\meter}$.  (d) Photograph of the IPT experimental setup using identical split-TLGRs made from aluminum.  The source and load coupling loops are suspended within the bores of the TLGRs. For scale, the transmitter and receiver are separated by a distance $x=\SI{10}{\centi\meter}$.}
\end{figure*}

This section describes the design of the LGRs used in the IPT systems that were investigated.  A basic CLGR consists of a conducting tube with a narrow slit along its length~\cite{Hardy:1981, Froncisz:1982}.  Shown in cross-section in Fig.~\ref{fig:LGRs}(a), this structure can be accurately modeled as a series $LRC$ circuit.  As shown in Fig.~\ref{fig:LGRs}(b), signals can be inductively coupled into and out of the LGR bore using coupling loops. The coupling loops are single-turn inductors that are made by short-circuiting the center conductor of a coaxial cable to its outer conductor~\cite{Rinard:1993}. Multi-turn inductive coupling loops can also be used when higher inductance is required.  The coupling strength between the coupling loop and nearby CLGR can be easily tuned by adjusting the distance between the two.

The dimensions of the CLGRs used in our work, given in Fig.~\ref{fig:LGRs}, were chosen such that $w\approx \left(\sqrt{2}-1\right)r_0$ which minimizes the resonant frequency $f_0$ for a fixed outer diameter \mbox{$d_\mathrm{c}=2\left(r_0+w\right)$}.  In the CLGR design, the quality factor is limited by losses due to radiated magnetic power.  Although radiative effects can be suppressed by surrounding the resonator by a concentric conducting cylinder~\cite{Hardy:1981, Bobowski:2013}, bulky shields can be impractical in many applications.  A TLGR can be made by joining two ends of a CLGR such that the bore of the resonator forms a torus.  The toroidal geometry confines the magnetic fields within the resonator bore such that high $Q$'s are achieved without requiring EM shields~\cite{Bobowski:2016}.  A schematic of the TLGR cross-section is shown in Fig.~\ref{fig:LGRs}(c).  The figure also shows a coupling loop suspended within the bore of the resonator.  In this case, the coupling strength is adjusted by changing the orientation of the loop relative to the bore cross-section.  The dimensions of the TLGR used in this work were chosen to satisfy $r_0\approx 5r_1/13$ which minimizes $f_0$ for a given resonator size characterized by \mbox{$d_\mathrm{t}=2\left(r_0+r_1\right)$}.  

Figure~\ref{fig:LGRs}(d) shows a pair of identical split-TLGRs used as a transmitter-receiver set in an IPT system.  Because currents loop around the axis of the toroidal bore, dividing the TLGR in half does not significantly alter the distribution or flow of charge and the resonant frequency is approximately unchanged.  In terms of an equivalent circuit model, dividing the resonator in half reduces the capacitance by a factor of two and increases the inductance by the same factor such that the resonant frequency \mbox{$\omega_0=2\pi f_0\approx 1/\sqrt{LC}$} remains unchanged.

One property that makes LGRs well-suited to IPT applications is their small electrical size. The resonant frequency of a CLGR is given by
\begin{equation}
f_0\approx \frac{1}{2\pi}\frac{c}{r_0}\sqrt{\frac{t}{\pi\varepsilon_\mathrm{r} w}},\label{eq:CLGRf0}
\end{equation}
where $c$ is the vacuum speed of light and $\varepsilon_\mathrm{r}$ is the dielectric constant of the material filling the gap of the resonator.  Using (\ref{eq:CLGRf0}) to solve for the free-space wavelength $\lambda=c/f_0$, setting \mbox{$w=\left(\sqrt{2}-1\right)r_0$}, and writing $w$ and $r_0$ in terms of the diameter $d_\mathrm{c}= 2(r_0 + w)$ leads to
\begin{equation}
\frac{d_\mathrm{c}}{\lambda}\approx 0.66\sqrt{\frac{t}{\varepsilon_\mathrm{r}d_\mathrm{c}}}.
\end{equation}
In our experiments, the resonator gap was filled with Teflon and $\varepsilon_\mathrm{r}\approx 2.1$.  Using the dimensions given in Fig.~\ref{fig:LGRs}(a), \mbox{$d_\mathrm{c}/\lambda\approx 0.017$} which is similar to the values achieved using helical and spiral resonators.    Manipulating any of $d_\mathrm{c}$, $\varepsilon_\mathrm{r}$, or $t$ to decrease $f_0$ will result in a decrease of the resonator size-to-wavelength ratio.  A similar analysis for the TLGR leads to
\begin{equation}
\frac{d_\mathrm{t}}{\lambda}\approx 2.48\sqrt{\frac{t}{\varepsilon_\mathrm{r}d_\mathrm{t}}}
\end{equation}
where $d_t = 2(r_0 + r_1)$.
Assuming a Teflon dielectric and uisng the dimensions given in Fig.~\ref{fig:LGRs}(c) results in $d_\mathrm{t}/\lambda\approx 0.048$. Although its size-to-wavelength ratio is several times larger than that of the CLGR, the toroidal geometry has other advantages, discussed below, that make it an attractive option for some IPT applications.

In the absence of radiative power losses, the unloaded quality factors of both the cylindrical and toroidal LGRs are approximately given by $Q_\delta\approx r_0/\delta$~\cite{Hardy:1981, Bobowski:2013, Bobowski:2016}.  For aluminum resonators with $f_0\approx\SI{100}{\mega\hertz}$, $r_0/\delta\sim 2000$.  As mentioned above, radiative effects in CLGRs are non-negligible and result in a suppression of the net \mbox{$Q=\left(Q_\delta^{-1}+Q_\mathrm{r}^{-1}\right)^{-1}$}, where $Q_\mathrm{r}$ is the quality factor associated with radiative losses.  The split-TLGRs used in our experiments, also  susceptible to radiative losses, had quality factors of $\sim 10^3$ which match the values typically obtained using helical and spiral resonators operating at $\sim\SI{10}{\mega\hertz}$.  It is worth noting that, for cylindrical and toroidal LGRs designed to have the minimum $f_0$ for a given $d$, \mbox{$Q_\delta\propto \left(td/\varepsilon_\mathrm{r}\right)^{1/4}$}.  To the extent that LGRs can be modeled as current loops, $Q_\mathrm{r}$ is proportional to $\left(\varepsilon_\mathrm{r}d/t\right)^{3/2}$~\cite{Snoke:2015}.  This analysis suggests that $Q_\mathrm{r}/Q_\delta=\left(\varepsilon_\mathrm{r}d^{5/7}/t\right)^{7/4}$ such that $Q_\mathrm{r}$ increases relative to $Q_\delta$ when the LGR resonant frequency is lowered by increasing $\varepsilon_\mathrm{r}$ or $d$, or by decreasing $t$.  Even if the current-loop model is not strictly correct, the insight that the radiative effects become less important as $f_0$ is decreased is still expected to apply.

In both the cylindrical and toroidal LGR designs, the electric fields are strongly confined to the gap of the resonator.  This is a key advantage for IPT applications because it makes the system insensitive to nearby dielectric objects, even if those objects are dynamic.  It also provides an opportunity to develop systems that wirelessly transfer power through a medium with non-negligible conductivity $\sigma$, such as saltwater~\cite{Wandinger:2021}.  Filling the gap with a low-loss dielectric, or otherwise isolating the gap from the conducting medium, eliminates the power dissipation (proportional to $\sigma |E|^2$) associated with the electric fields $\bm{E}$. 

The bore of CLGRs, modeled as a single-turn inductor, produce magnetic fields that form closed loops by exiting one end of the bore and entering the opposite end.  In this way, a pair of coupled CLGRs have magnetic field configurations that are qualitatively similar to those obtained in IPT systems using conventional helical and spiral resonators. Therefore, a IPT system implemented with CLGRs is expected to have similar spatial sensitivity relative to other systems implemented using helical or spiral resonators~\cite{Sample:2013}. 

In applications that do not require omnidirectional power transfer, such as when the lateral alignment of transmitter and receiver is fixed, the split-TLGRs offer another unique advantage.  The geometry ensures that, for a pair of coupled split-TLGRs, the magnetic flux is very weak everywhere except within the bores of the resonators and in the space directly between transmitter and receiver.  A detailed analysis of the EM field configurations of LGR IPT systems is presented next. 

\section{\label{sec:sims}Electromagnetic Simulations}
\begin{figure*}
\begin{tabular}{c}
(a)~\includegraphics[width=1.4\columnwidth]{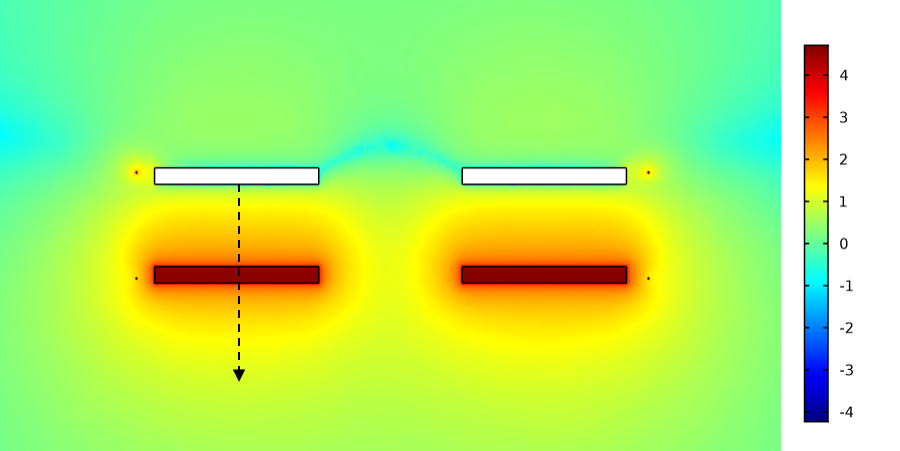}\\
~\\
(b)~\includegraphics[width=1.4\columnwidth]{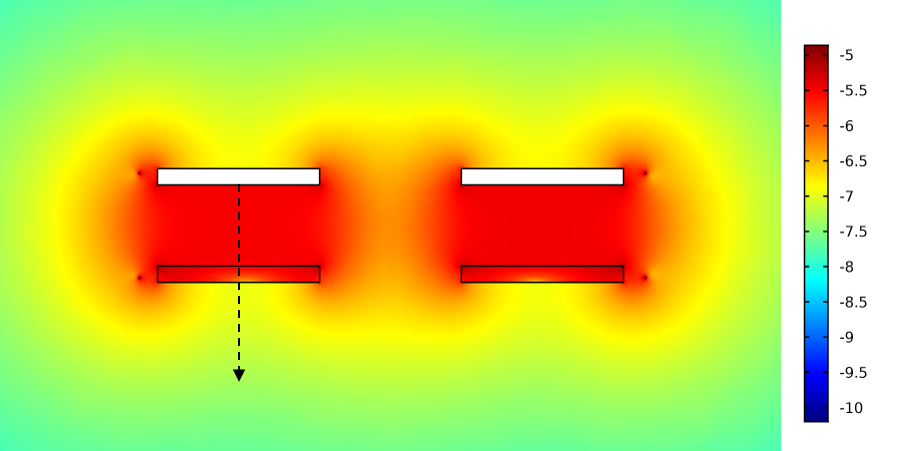}\\
 ~\\
\begin{tabular}{cc}
(c)~\includegraphics[width=0.92\columnwidth]{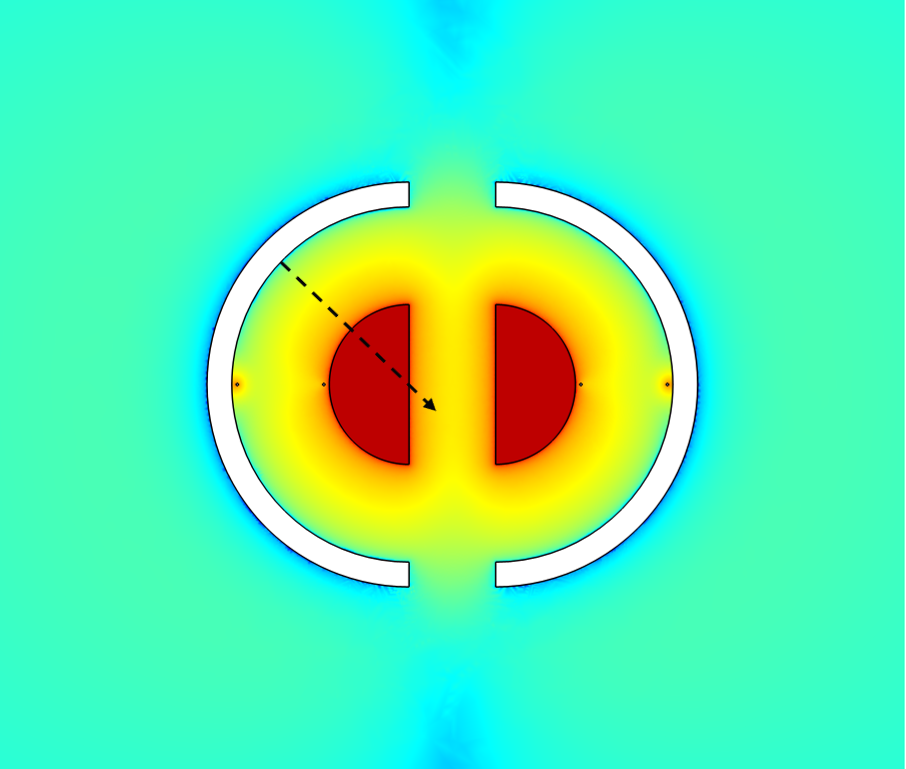} & (d)~\includegraphics[width=0.92\columnwidth]{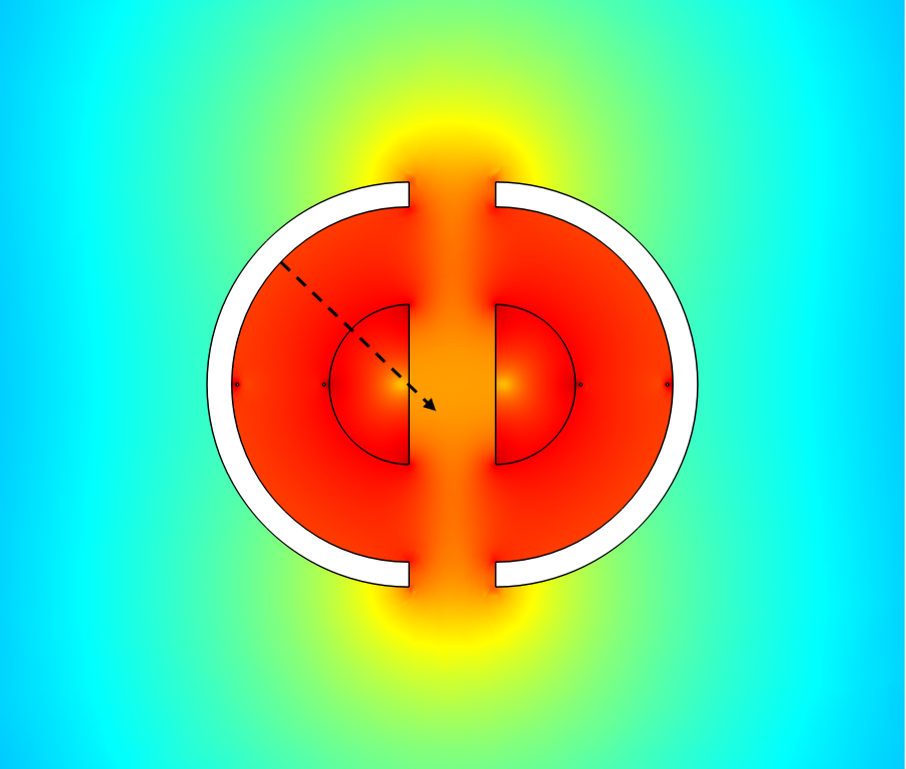}
\end{tabular}
\end{tabular}
\caption{\label{fig:COMSOLEM}The simulated (a) electric and (b) magnetic field magnitudes for a CLGR-based IPT system.  The color scales are $\log E$ and $\log B$, respectively.  The dashed arrows show the cut lines that were used to generate the plots in Fig.~\ref{fig:COMSOLscan}(a).  The simulated (c) electric and (d) magnetic field magnitudes for a TLGR-based IPT system shown using logarithmic color scales.  The color scales of (c) and (d) are identical to the scales used in (a) and (b), respectively.  The dashed arrows show the cut lines that were used to generate the plots in Fig.~\ref{fig:COMSOLscan}(b).
}
\end{figure*}

\begin{figure*}
\begin{tabular}{cc}
(a)~\includegraphics[width=0.9\columnwidth]{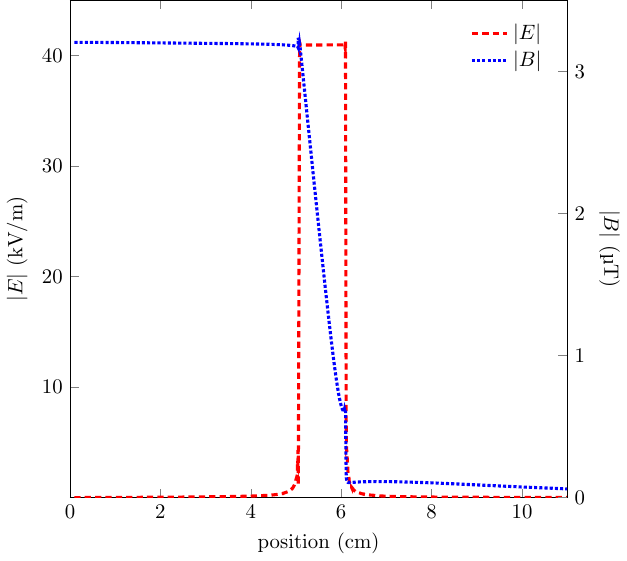} & (b)~\includegraphics[width=0.9\columnwidth]{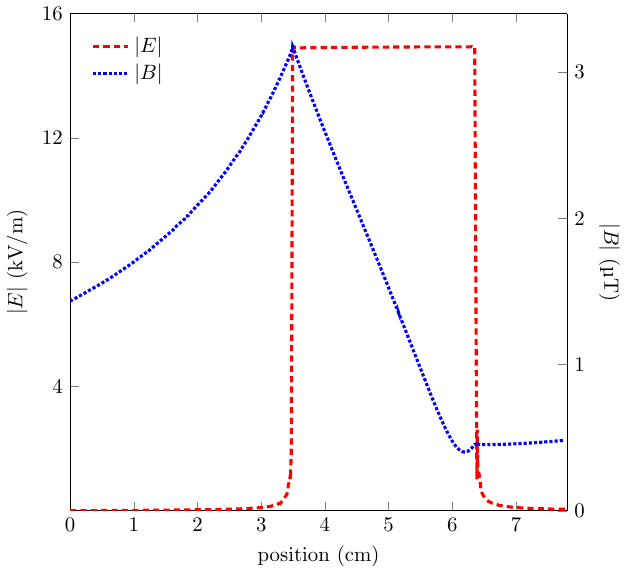}
\end{tabular}
\caption{\label{fig:COMSOLscan}The electric and magnetic field strengths as a function of position through the resonator bore and gap for coupled (a) CLGRs and (b) split-TLGRs.  The paths followed to generate these plots are shown in Figs.~\ref{fig:COMSOLEM}(a)--(d).  For both the CLGR and TLGR IPT systems, the electric field strength is very weak outside of the gap region.  For the CLGR, the magnetic field strength is nearly constant through the bore and then decreases linearly through the gap.  In the case of the TLGR, the magnetic field has a $1/r$ dependence through the bore and then falls off linearly through the gap. 
}
\end{figure*}

\begin{figure*}
\begin{tabular}{cc}
(a)~\includegraphics[height=5.05cm]{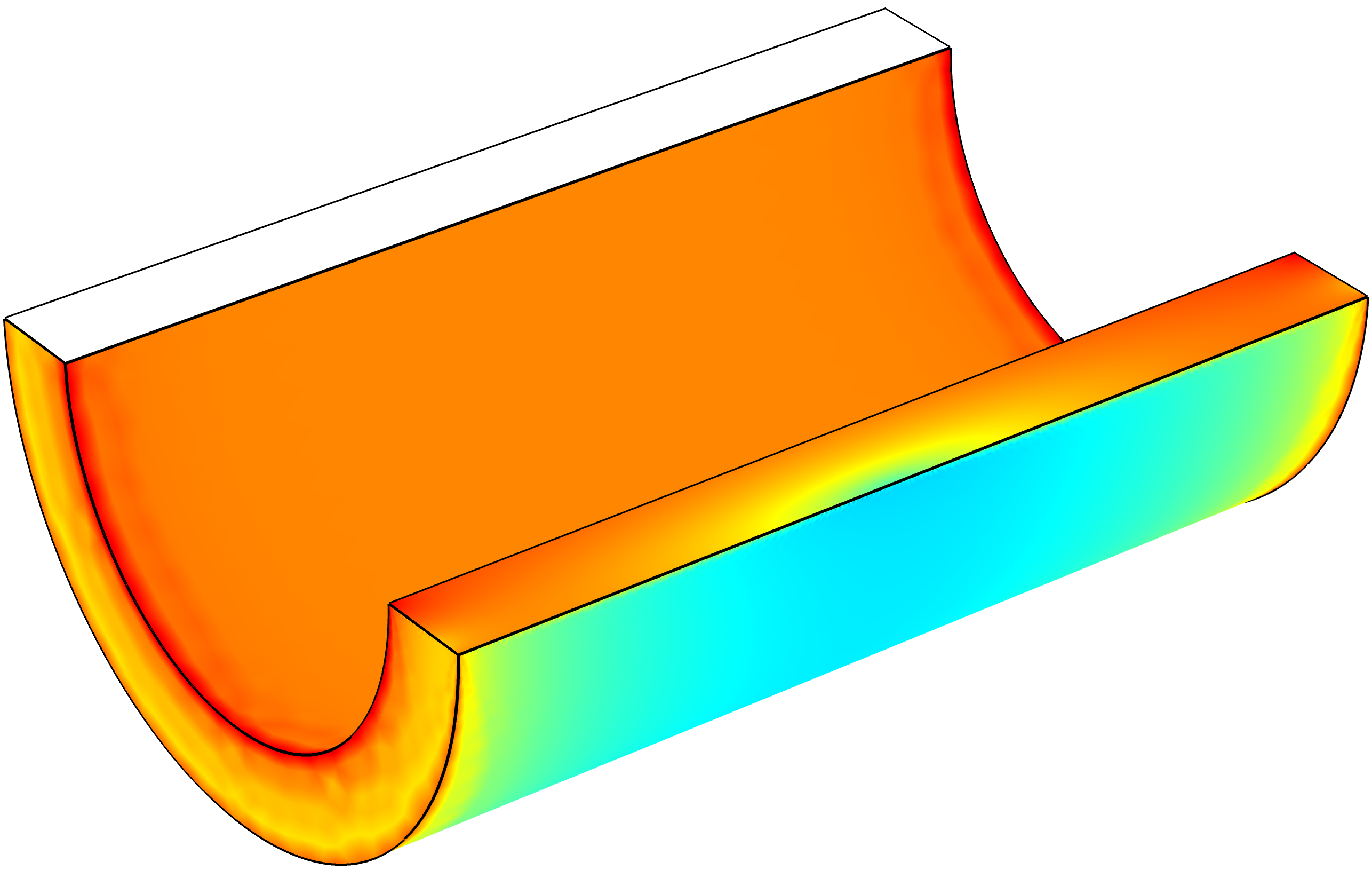} & (b)~\includegraphics[height=6.1cm]{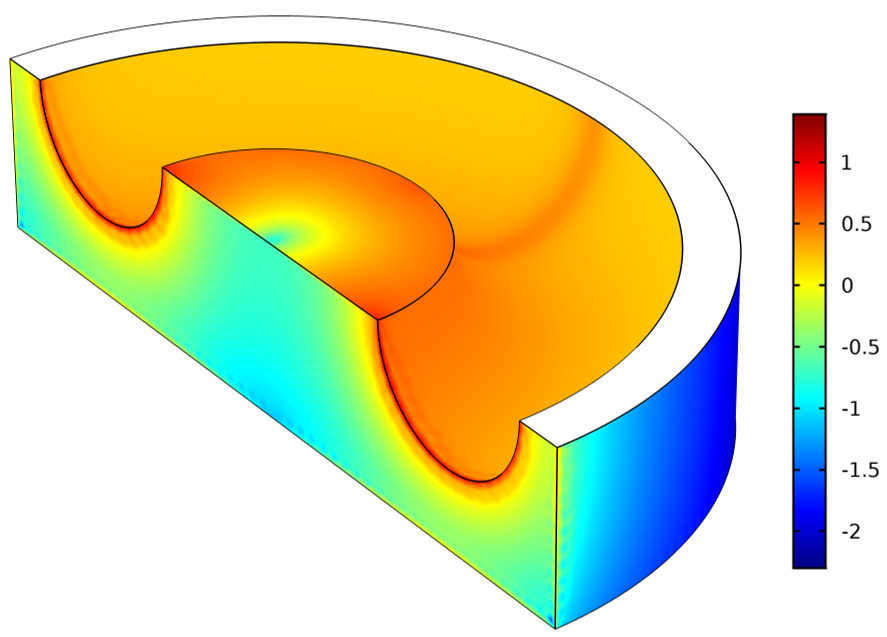}
\end{tabular}
\caption{\label{fig:currentDensity}Surface current density maps for the (a) CLGR and (b) TLGR IPT systems.  The color scales represent $\log J_\mathrm{s}$ and are identical for (a) and (b).  For clarity, only the bottom halves of the transmitting resonators are shown.  The white surfaces represent cuts through solid aluminum.
}
\end{figure*}

Numerical simulations using a finite element solver in COMSOL were used to investigate the EM fields in LGR IPT systems.  Complete models of the four-coil IPT systems were created with \SI{50}{\ohm} ports exciting a narrow gap in the source and load coupling loops.  The amplitude of the source delivers a peak voltage of \SI{1}{\volt} to a matched port. The entire model was surrounded by a spherical boundary with a perfectly matched layer.  In all simulations, we confirmed that the spherical boundary had negligible effects on the results.

The aluminum resonators were modeled with a finite resistivity of \SI{2.65e-8}{\ohm\meter}.  The capacitive gap in the model was set to \SI{102}{\micro\meter} to match the design value and it was filled with a lossless dielectric.  Because it was difficult to accurately measure the gap size of the experimental resonators, the simulated resonant frequencies were tuned to match the experimental values by adjusting the dielectric constant of the gap material.

A simulation was set up to model an IPT link with CLGRs. The distance $x$ between the transmit and  receive CLGRs was \SI{88.8}{\centi\meter} and the dielectric constant in the gap region was set to $\varepsilon_\mathrm{r}=1.4$. The coupling loop positions were tuned to achieve critical coupling and the simulation was run at the frequency of optimum power transfer efficiency.  The coupling loop diameters were \SI{65.4}{\milli\meter}, the source loop-transmitter distance was \SI{10.75}{\milli\meter}, the load loop-receiver distance was \SI{13.2}{\milli\meter}, and the frequency was $\SI{102.1}{\mega\hertz}$.  

The simulation results for the CLGR IPT system are shown in Figs.~\ref{fig:COMSOLEM}(a) and (b). In
Fig.~\ref{fig:COMSOLEM}(a), the magnitude of the electric field $E$ is shown for a plane that passes through the centers of the CLGR gaps.  The corresponding  cross-section is shown as section $AA$  in Fig.~\ref{fig:LGRs}(a).  The color scale in Fig.~\ref{fig:COMSOLEM}(a)  is  logarithmic  to illustrate how rapidly the electric field strength decreases as one moves away from the capacitive gap. Figure~\ref{fig:COMSOLEM}(b) shows the corresponding magnitude of the magnetic field $B$ for the CLGR IPT. The color scale is logarithmic and the magnetic field is strongest within the bores of the CLGRs. Outside the CLGRs, the magnetic field is weaker and extends to the surrounding space to provide magnetic coupling between the transmitter and receiver. Also note that the magnetic field extends through the capacitive gap  region in the CLGRs. The gap region is at the bottom of Fig.~\ref{fig:COMSOLEM}(b) and the field decreases across the gap as the outer edge is approached.

Figures~\ref{fig:COMSOLEM}(c) and (d) show the simulated electric and magnetic field strengths for the split-TLGR IPT system.  The fields are shown through section $BB$ indicated in Fig.~\ref{fig:LGRs}(c). So that direct comparisons between the CLGR and TLGR systems can be made, Figs.~\ref{fig:COMSOLEM}(a) and (c) and Figs.~\ref{fig:COMSOLEM}(b) and (d) use the same color scales.  In the TLGR case, critical coupling and optimum power transfer were achieved using $x=\SI{31}{\centi\meter}$, $\varepsilon_\mathrm{r}=2.1$, and $f=\SI{122.5}{\mega\hertz}$.  The coupling loops had a radius of \SI{15.5}{\milli\meter} and the planes of the loops were aligned with the cross-sections of the split-TLGR bores.  Once again, the electric field is weak everywhere outside the gap. The corresponding magnetic field for the TLGR IPT system is shown in Fig.~\ref{fig:COMSOLEM}(d). Unlike the CLGR case, the magnetic field is primarily confined to the direct bore coupling region and rapidly decreases outside this region.  The magnetic field strength in the region behind the split-TLGRs is approximately two orders of magnitude less than that in the space linking the bores of the pair of resonators. It is this property that makes the split-TLGRs unique among all of the other resonators currently used in IPT systems.

For the coupled CLGRs, Fig.~\ref{fig:COMSOLscan}(a) shows plots of $E$ and $B$ as a function of position along the dashed lines shown in Figs.~\ref{fig:COMSOLEM}(a) and (b).  The electric field a distance $w=\SI{1.02}{\centi\meter}$ from the outside edge of the gap is more than \num{2.5} orders of magnitude less than the peak value of $E$.  The magnetic field is uniform within the bore and then linearly decreases through the gap, a behavior that was first predicted by the theoretical analysis of Hardy and Whitehead in \cite{Hardy:1981}.  These results show that the simulations provide reliable solutions for the EM fields of the IPT models.  Figure~\ref{fig:COMSOLscan}(b) shows the corresponding $E$ and $B$ plots for the TLGR system along the dashed lines of Figs.~\ref{fig:COMSOLEM}(c) and (d).  In this case, a diagonal path was chosen to avoid extrinsic effects due to the coupling loop and the ends of the split-TLGR bore.  Once again, the electric field strength is negligible outside the gap.  The magnetic field strength has a linear dependence through the gap and then falls off as $1/r$ through the bore which is characteristic of the magnetic field of a toroid.  

An important observation is that the substantial magnetic flux contained within the gap of the TLGR limits the maximum possible coupling to the loop suspended within the bore.  Specifically, the coupling coefficients $k_\mathrm{t}$ and $k_\mathrm{r}$ are expected to be significantly less than one, even with the plane of the loop set perpendicular to the axis of the toroidal bore (i.e.\ maximum coupling).

Figure~\ref{fig:currentDensity} shows the magnitudes of the simulated surface current densities $\bm{J}_\mathrm{s}$ on a logarithmic scale for both the CLGR and TLGR IPT systems.  These figures were generated from the same simulations used to create Figs.~\ref{fig:COMSOLEM} and \ref{fig:COMSOLscan}.  For clarity, in both (a) and (b), only the bottom halves of the transmitting resonators are shown.  Identical color scales are used for the pair of images.  As expected, the surface current densities are nearly uniform along the inner surfaces of the LGR bores.  Figure~\ref{fig:currentDensity}(a) shows that $J_\mathrm{s}$ is enhanced at ends of the CLGR bore which is where local hot spots due to Joule heating would be expected in high-power applications.  For the TLGR, Fig.~\ref{fig:currentDensity}(b) shows that $J_\mathrm{s}$ is again largest at the ends of the bore.  The surface current density is also enhanced along the outer diameter of the TLGR gap and near the midpoint of the bore where the coupling loop is suspended (not shown in the figure).

\section{\label{sec:expt}Experimental Results}
\begin{figure}
\includegraphics[height=8cm]{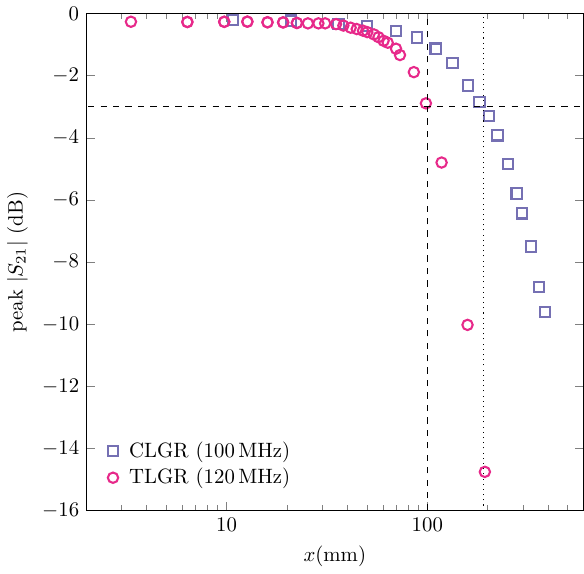}
\caption{\label{fig:peakS21}Peak $\left\vert S_{21}\right\vert$ as a function of transmitter-receiver separation distance $x$ for the CLGR and TLGR IPT systems.  The horizontal dashed line is the \SI{-3}{\decibel} point and is used to define the spatial bandwidth $x_0$ of the IPT systems.  The vertical dashed and dotted lines indicate the spatial bandwidths of the TLGR and CLGR systems, respectively. 
}
\end{figure}

\subsection{\label{sub:exptCopmpare}Power Transfer Efficiency Versus Distance}
This section begins with an experimental comparison of the power transfer efficiency of CLGR- and TLGR-based IPT systems.  First, the transmitting and receiving LGRs were set to have identical resonant frequencies such that \mbox{$f_\mathrm{t}=f_\mathrm{r}\equiv f_0$}. For each resonator, the frequency was set by adjusting the fraction of the gap that was filled with Teflon. After tuning, the resonant frequencies for the CLGR and split-TLGR systems were \SI{100}{\mega\hertz} and \SI{120}{\mega\hertz}, respectively.  Next, the distance $x$ between the transmitter and receiver was set and the couplings $k_\mathrm{t}$ and $k_\mathrm{r}$ were adjusted for maximum power transfer efficiency. 

Figure~\ref{fig:peakS21} shows plots of the peak $\left\vert S_{21}\right\vert$ of the tuned systems as a function of separation distance $x$.  It is important to emphasize that the peak $\left\vert S_{21}\right\vert$ frequencies are different at each value of $x$ in Fig.~\ref{fig:peakS21}.  Therefore, the performance suggested in the figure could only be achieved while employing a frequency tracking system.  We consider fixed-frequency operation in Section~\ref{sub:bandwidth}.  At the smallest values of $x$, the CLGR system achieves a peak $\left\vert S_{21}\right\vert\approx \SI{-0.21}{\decibel}$ which corresponds to a power transfer efficiency of \SI{95}{\percent}.  The efficiency remains high as $x$ is increased and then rapidly drops off after passing a critical distance $x_0$ which we define to be the \SI{-3}{\decibel} point.  We refer to $x_0$ as the spatial bandwidth and, for the CLGR system, it was found to be $x_0\approx\SI{19.0}{\centi\meter}$.  The roll off above $x_0$ is steep with $\left\vert S_{21}\right\vert$ dropping by approximately \SI{7}{\decibel} when $x$ doubles from \num{20} to \SI{40}{\centi\meter}.  

The peak $\left\vert S_{21}\right\vert$ versus $x$ measurements for the TLGR system follow the same general trends found with the CLGRs.  At the smallest values of $x$, $\left\vert S_{21}\right\vert$ peaked at \SI{-0.26}{\decibel} (\SI{94}{\percent} efficiency).  The spatial bandwidth of this system was \mbox{$x_0=\SI{10.0}{\centi\meter}$} and the roll off for $x>x_0$ was very steep, dropping by approximately \SI{13.0}{\decibel} when the distance was doubled from \num{10} to \SI{20}{\centi\meter}.

The results from Fig.~\ref{fig:peakS21} clearly suggest that the CLGR system outperforms the TLGR system. It has a higher spatial bandwidth and the roll off above $x_0$ is slower.  A useful figure of merit for IPT systems is $x_0/s$ where $s$ characterizes the largest dimension of the transmitter/receiver.  We use $s_\mathrm{c}$ and $s_\mathrm{t}$ to represent the largest dimensions of the CLGR and TLGR, respectively.  For the CLGRs, \mbox{$s_\mathrm{c}=\SI{10.2}{\centi\meter}$} is set by the resonator length such that \mbox{$x_0/s_\mathrm{c}=1.9$.}  This value of $s_\mathrm{c}$ does not include the space between the coupling loop and the resonator.  However, as was done with the TLGR, this space could be eliminated by mounting the coupling loop within the bore the resonator.  The geometry of the split-TLGRs makes \mbox{$s_\mathrm{t}=\SI{12.7}{\centi\meter}$} relatively large such that \mbox{$x_0/s_\mathrm{t}=0.79$} is comparatively small.  

Table~\ref{tab:Tab1} compares the figures of merit of various IPT systems reported in the literature. 
\begin{table}
\caption{\label{tab:Tab1}Figure of merit comparison of IPT systems in the published literature. The parameter $x_0$ is the spatial bandwidth of the IPT system and $s$ characterizes the size of the transmitter/receiver}
\centering
\begin{tabular}{cccccc}
\hline\hline\\[-1.5ex]
Resonator\\ Type & $f_0$ (\si{\mega\hertz}) & $s$ (\si{\centi\meter}) & $x_0$ (\si{\centi\meter}) & $x_0/s$ & Reference\\[0.5ex]
\hline\\[-1.5ex]
CLGR & \num{100} & \num{10.2} & \num{19} & \num{1.9} & --\\
TLGR & \num{120} & \num{12.7} & \num{10} & \num{0.79} & --\\
helical & \num{10.56} & \num{60} & \num{180} & \num{3.0} & \cite{Soljacic:2007}\\
coaxial & \num{30.8} & \num{20} & $<\num{60}$ & $<\num{3.0}$ & \cite{Dionigi:2012}\\
dielectric & \num{232} & \num{8.4} & \num{16} & \num{1.9} & \cite{Song:2016}\\
spiral & \num{7.65} & \num{59} & \num{100} & \num{1.7} & \cite{Sample:2011}\\
spiral & \num{6.7} & \num{56} & \num{85} & \num{1.5} & \cite{Duong:2011}\\
helical & \num{16.1} & \num{15} & \num{15} & \num{1.0} & \cite{Cheon:2011}\\
coil & \num{0.742} & \num{3.8} & \num{2.5} & \num{0.66} & \cite{Li:2012}\\
coil & \num{0.7} & \num{6.4} & \num{4.0} & \num{0.63} & \cite{RamRakhyani:2011}\\[0.5ex]
\hline\hline
\end{tabular}
\end{table}
The first observation is that the CLGR system has a figure of merit that is competitive with many of the systems reported in the literature.  We note that, because the CLGR resonant frequency is independent of the resonator's length $\ell$, it may be possible to increase its figure of merit by decreasing $\ell$. An additional increase in $x_0/s_\mathrm{c}$ may be expected if the operating frequency could be decreased from \SI{100}{\mega\hertz} to a value closer to \SI{10}{\mega\hertz} that is more typical of IPT systems. 

A more striking observation from Table~\ref{tab:Tab1} is that, despite more than a decade of intense research since 2007, the IPT system with the best figure of merit is still the original system reported by the Solja\v{c}i\'{c} group~\cite{Soljacic:2007}.  This realization motivated us to shift our focus away from improving $x_0/s$ and towards designing a resonator with desirable attributes not found among the transmitters and receivers currently used in IPT systems.

Specifically, our focus has been on shaping the electric and magnetic field configurations in and around the transmitter and receiver.  As shown earlier in Figs.~\ref{fig:COMSOLEM}(a) and (c), the electric fields in both the CLGR and TLGR systems are strongly confined to the gap of the resonators. Furthermore, Fig.~\ref{fig:COMSOLEM}(d) shows that the magnetic field external to the TLGRs is predominantly within the region of space between the transmitter and receiver.  Anecdotally, when measuring $\left\vert S_{21}\right\vert$ as a function of $x$, we found the CLGR system to be much more sensitive to its surrounding environment.  To acquire reliable data, it was necessary to move away from the experimental apparatus before recording the $\left\vert S_{21}\right\vert$ traces.  The peak value of $\left\vert S_{21}\right\vert$ changed by several tenths of a decibel when the system was approached to adjust either the coupling loop positions or $x$.  In contrast, the split-TLGR system did not exhibit a sensitivity to similar changes to its surroundings. Therefore, although the CLGR system has more compact resonators, the split-TLGR has the advantage that it is more robust to changing environmental conditions in the neighboring power transfer space.

\begin{figure*}
\begin{tabular}{cc}
(a)~\includegraphics[height=8cm]{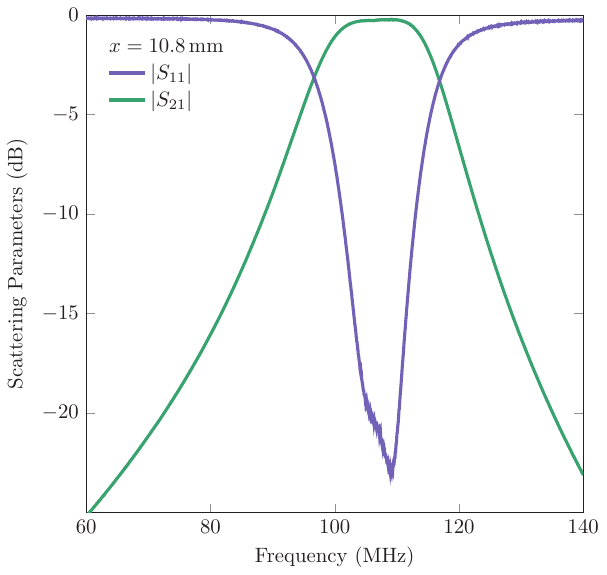} & (b)~\includegraphics[height=8cm]{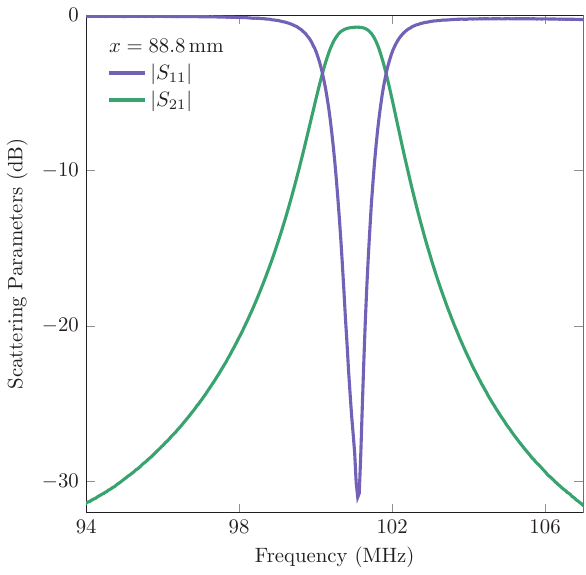}\\
~ & ~\\
(c)~\includegraphics[height=8cm]{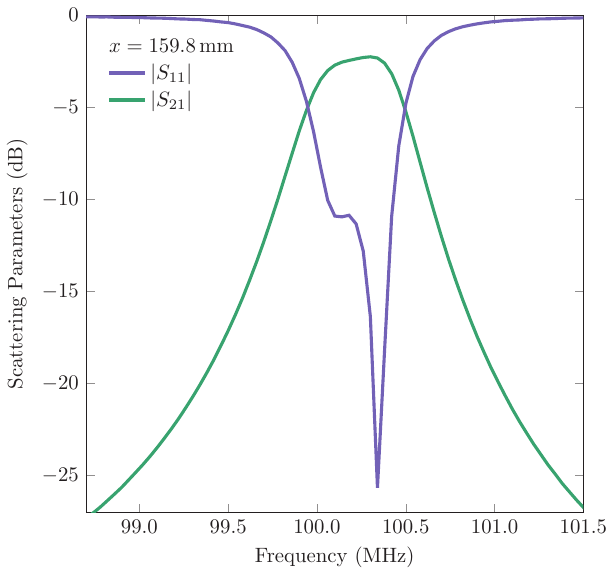} & (d)~\includegraphics[height=8cm]{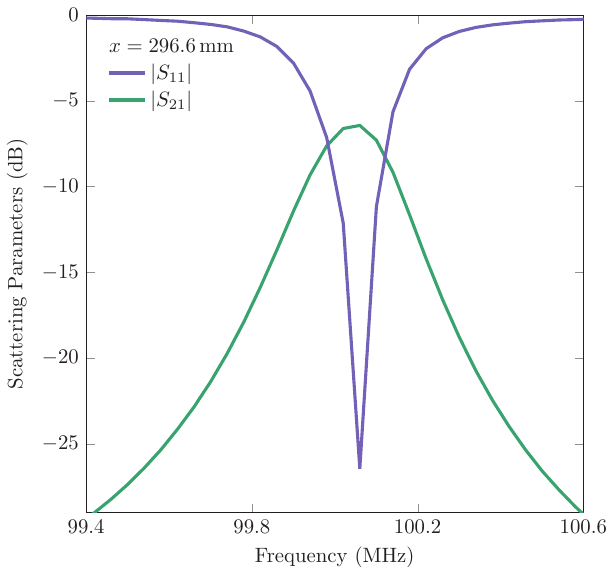}
\end{tabular}
\caption{\label{fig:S21CLGR}The $\left\vert S_{11}\right\vert$ and $\left\vert S_{21}\right\vert$ responses for the CLGR IPT system for different  separation distances.  In (a) and (b), $x<x_0$, and the system is critically coupled.  In (c), $x$ is close to the -3~dB spatial bandwidth $x_0$ and the transmission efficiency starts to decrease.  In (d), $x>x_0$; the system is undercoupled and significant attenuation is shown in the $\left\vert S_{21}\right\vert$ response.  Note that for all four conditions, the IPT system is matched and the corresponding $\left\vert S_{11}\right\vert$ is typically better than \SI{-25}{\decibel} at the resonant frequencies. Also, note the different frequency ranges used in the four plots.}
\end{figure*}

\begin{figure*}
\begin{tabular}{cc}
(a)~\includegraphics[height=8cm]{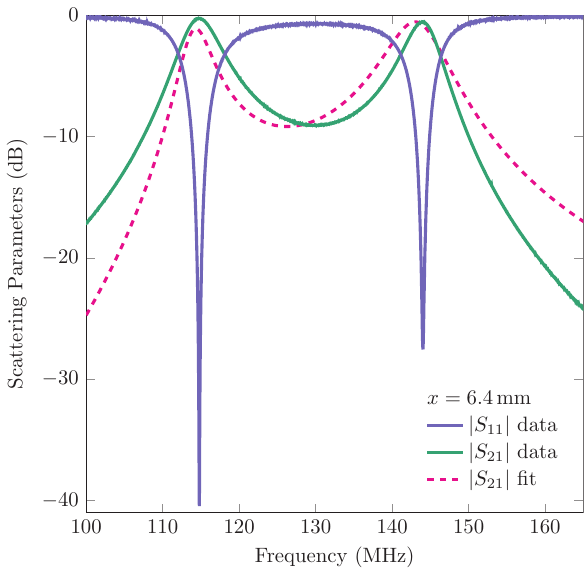} & (b)~\includegraphics[height=8cm]{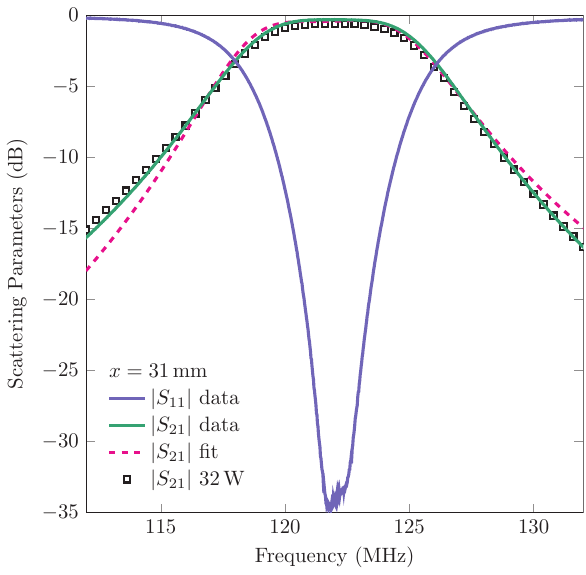}\\
~ & ~\\
(c)~\includegraphics[height=8cm]{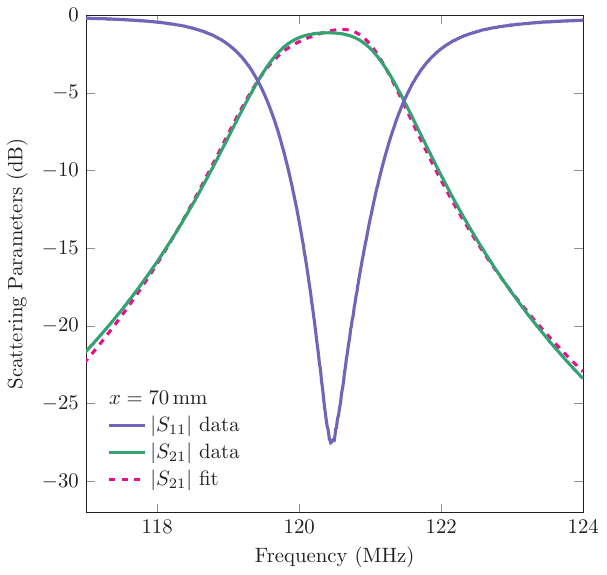} & (d)~\includegraphics[height=8cm]{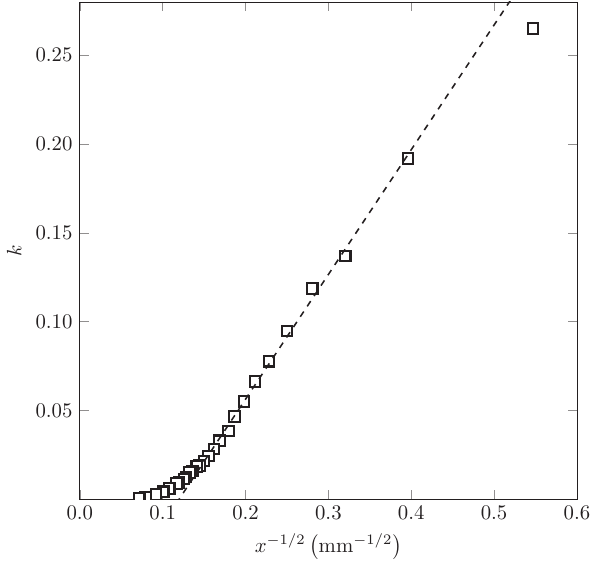}
\end{tabular}
\caption{\label{fig:S21}Fits to the frequency dependence of $\left\vert S_{21}\right\vert$ for the system of coupled split-TLGRs at various values of the separation distance $x$.  The equivalent circuit of Fig.~\ref{fig:circuits}(b) was used to calculate the theoretical $\left\vert S_{21}\right\vert$ frequency dependence and the fits are used to extract experimental values of the the coupling coefficient $k$.  Although not used in the analysis, for completeness, the $\left\vert S_{11}\right\vert$ response are also shown.  (a) In the $x=\SI{6.4}{\milli\meter}$ data, the IPT system is overcoupled and $\left\vert S_{21}\right\vert$ exhibits a double resonance.  (b) At $x=\SI{31}{\milli\meter}$, $\left\vert S_{21}\right\vert$ has a broad and flat peak and the power transfer efficiency is high.  The open square data show a measurement of $\left\vert S_{21}\right\vert$, made using directional couplers and power sensors, when the IPT system was operated at \SI{32}{\watt}.  (c) At $x=\SI{70}{\milli\meter}$, the $\left\vert S_{21}\right\vert$ peak has narrowed and dropped in magnitude.  Note that at all values of $x$, $\left\vert S_{11}\right\vert$ is typically better than \SI{-30}{\decibel} at the resonant frequencies.  Also, note the differences in the frequency range used for the plots in (a) -- (c).    (d) The values of $k$ extracted from the fits as a function of $x^{-1/2}$.  For small values of $x$, $k$ approximately follows a $x^{-1/2}$ dependence (dashed line).
}
\end{figure*}

\begin{figure*}
\begin{tabular}{cc}
(a)~\includegraphics[height=8cm]{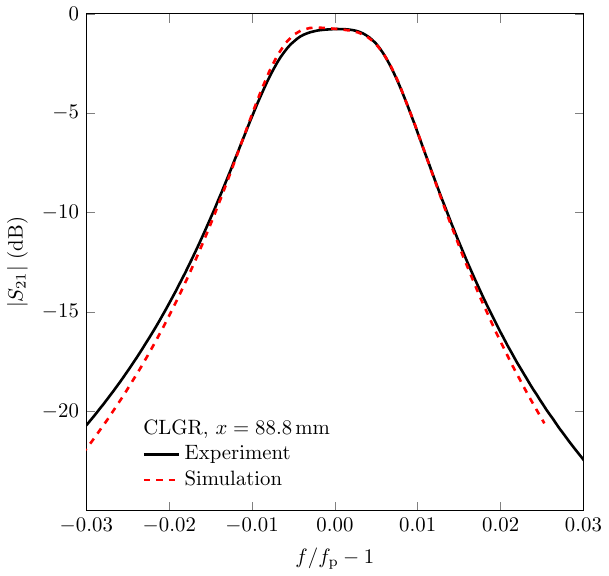} & (b)~\includegraphics[height=8cm]{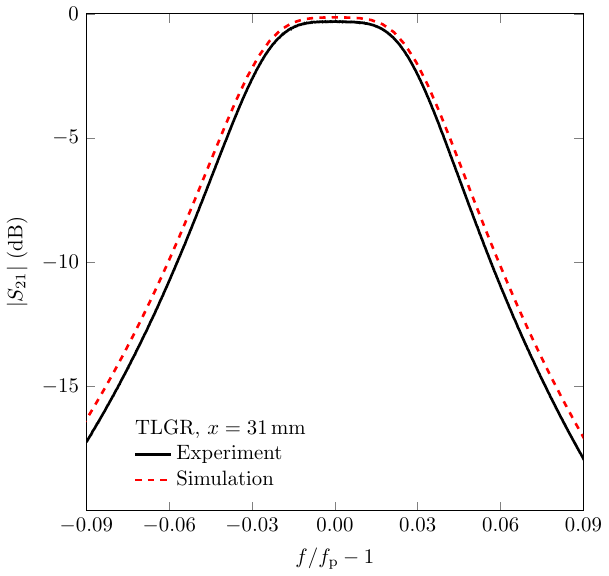}
\end{tabular}
\caption{\label{fig:ExptSim}Comparison of $\left\vert S_{21}\right\vert$ measured experimentally and extracted from COMSOL simulations. (a) The CLGR IPT system when $x=\SI{88.8}{\milli\meter}$.  (b) The TLGR IPT system when $x=\SI{31}{\milli\meter}$.  For comparison purposes, horizontal axes of $f/f_\mathrm{p}-1$ were chosen so as to place the center of the $\left\vert S_{21}\right\vert$ peaks at zero.}
\end{figure*}

\subsection{\label{sub:expt1}LGR IPT System Characterization}
We now present a detailed experimental characterization of the LGR IPT systems.  For each transmitter-receiver separation distance $x$, the source and load coupling loop positions/orientations were tuned to achieve optimal power transfer efficiency.  Figures~\ref{fig:S21CLGR} and \ref{fig:S21} show example $\left\vert S_{11}\right\vert$ and $\left\vert S_{21}\right\vert$ responses measured at different values of $x$ for the CLGR and split-TLGR systems, respectively.  Although not shown, we also recorded $\left\vert S_{22}\right\vert$ and $\left\vert S_{12}\right\vert$ traces at each value of $x$.  As expected, interchanging the transmitting and receiving resonators had no significant effect on the measured reflection and transmission coefficients.  

At the smallest values of $x$, the coupling $k$ between the transmitting and receiving resonators is strong.  To achieve optimal power transfer at a single mode under these conditions, the coupling coefficients $k_\mathrm{t}$ and $k_\mathrm{r}$ are also required to be large.  In the case of the split-TLGRs, we found that for $x$ less than \SI{31}{\milli\meter}, even with the planes of the coupling loops aligned perpendicular to the axis of the bores, we could not obtain $k_\mathrm{t}$ and $k_\mathrm{r}$ values large enough to observe efficient single-mode power transfer.  In this case, the IPT system is said to be overcoupled and exhibits a double resonance~\cite{Sample:2011}.  

Figure~\ref{fig:S21}(a) is an example of an overcoupled TLGR IPT system at $x=\SI{6.4}{\milli\meter}$.  Efficient power transfer is still possible, however, the frequencies of the two normal modes occur above and below the natural resonance frequency of the TLGRs.  The low-frequency peak, in which the currents in the two resonators are in phase, is called the even mode while, for the high-frequency odd mode, the currents are out of phase.  Note that the peak height of the even mode is slightly higher than that of the odd mode.  This phenomenon can be understood in terms of the cross-couplings.  In the even mode, the small amount of magnetic flux from the transmitter that couples directly to the load loop adds constructively to the primary contribution from the receiver.  In the odd mode,  the two sources of flux add destructively~\cite{Sample:2011}.  

We note that it would be possible to reach smaller values of $x$ without entering into the overcoupled regime by increasing the values of $L_\mathrm{s}$ and $L_\ell$.  This could be done by constructing coupling loops with two or more loops of wire.  In the case of the CLGR system, the the coupling loops were outside the bore the resonator such that larger-diameter loops could be used.  For this reason, the overcoupled regime was avoided in this system, even at the smallest values of $x$.  

When coupling coefficients are tuned for a single mode and $x$ is less than the spatial bandwidth $x_0$, the IPT system is critically coupled.  In this case, the $\left\vert S_{21}\right\vert$ frequency response has a single resonance with a broad, flat peak.  Figures~\ref{fig:S21CLGR}(a) and (b) show $\left\vert S_{21}\right\vert$ for a critically coupled  CLGR system when $x=\SI{10.8}{\milli\meter}$ and \SI{88.8}{\milli\meter}, respectively.  Figure~\ref{fig:S21}(b) shows a critically coupled split-TLGR system when $x=\SI{31}{\milli\meter}$.  The figures also show the measured $\left\vert S_{11}\right\vert$ responses.  The minimum $\left\vert S_{11}\right\vert$ values at the resonant frequencies are  $<\SI{-20}{\decibel}$ indicating that very little power is reflected back to the source.  

As $x$ is increased and begins to approach the spatial bandwidth $x_0$, the coupling $k$ between the transmitter and receiver decreases and $k_\mathrm{t}$ and $k_\mathrm{r}$ likewise have to be decreased to maintain optimal power transfer.  In this regime, the widths of the $\left\vert S_{21}\right\vert$ resonances narrow and the peak values drop as $x$ is increased.  Figures~\ref{fig:S21CLGR}(c) and \ref{fig:S21}(c) show examples of this response for the CLGR and TLGR IPT systems.  Finally, for $x>x_0$, the power transfer efficiency drops rapidly and the IPT system is said to be undercoupled.  Figure~\ref{fig:S21CLGR}(d) shows $\left\vert S_{21}\right\vert$ for an undercoupled CLGR system. 

\subsection{\label{sub:S21ExptvsSim}Comparison of Experimental and Simulation Results}
Section~\ref{sec:sims} showed EM field simulation results for the IPT systems. The simulations were also used to extract the frequency dependencies of the scattering parameters for various transmitter-receiver separation distances.  Figure~\ref{fig:ExptSim} compares the experimental and simulated $\left\vert S_{21}\right\vert$ responses.  Due to small but unavoidable differences in geometry, the resonant frequencies of the physical LGRs and 3-D models constructed in COMSOL were not identical.  Therefore, in Fig.~\ref{fig:ExptSim} we have plotted the transmission coefficient data as a function of $f/f_\mathrm{p}-1$, where $f_\mathrm{p}$ represents the center frequency of the $\left\vert S_{21}\right\vert$ peak.  

Figure~\ref{fig:ExptSim}(a) compares $\left\vert S_{21}\right\vert$ responses for the CLGR system when $x=\SI{88.8}{\milli\meter}$.  The experimental data is identical to the data shown in Fig.~\ref{fig:S21CLGR}(b) and has been plotted using \mbox{$f_\mathrm{p,e}=\SI{101.04}{\mega\hertz}$}.  The corresponding simulated transmission coefficient was plotted using $f_\mathrm{p,s}=\SI{102.41}{\mega\hertz}$.  The agreement between the experimental and simulated results is remarkable.  Figure~\ref{fig:ExptSim}(b) compares the $\left\vert S_{21}\right\vert$ frequency sweeps for the TLGR IPT geometry when $x=\SI{31}{\milli\meter}$ and using $f_\mathrm{p,e}=\SI{121.96}{\mega\hertz}$ and $f_\mathrm{p,s}=\SI{122.45}{\mega\hertz}$.  The agreement is once again very good, with the simulated system exhibiting slightly less loss at all frequencies.

There are several possible explanations for the small differences observed between the experimental and simulated transmission coefficients.  First, in the COMSOL models, it is difficult to precisely replicate the experimental conditions that set the values of $k_\mathrm{t}$ and $k_\mathrm{r}$.  The simulated coupling loop is perfectly circular and planar whereas the physical structure was handmade and has imperfections.  Moreover, one cannot achieve perfect lateral and planar alignment of the coupling loops with the CLGRs on the testbed.  In the case of the TLGRs, neither the centering of the coupling loops within the resonator bore nor the setting of the orientations can be done precisely on the testbed.  A second consideration is that the simulated scattering parameters were extracted without taking into account losses that occur within the lengths of UT-141 semi-rigid coaxial cable that lead up to the coupling loops.  These sections of cable are approximately \SI{15}{\centi\meter} long and are terminated with an SMA connector.  In the experimental data, the VNA calibration plane is established at the SMA connector and losses that occur along the length of the coaxial cable leading up to the coupling loop are included in the $\left\vert S_{21}\right\vert$ measurements.  Finally, the COMSOL model assumed an aluminum resistivity of $\rho=\SI{2.65e-8}{\ohm\meter}$ and a lossless dielectric filling the resonator gap. Although these assumptions are reasonable, a small discrepancy in the resistivity and a non-negligible loss tangent in the Teflon dielectric used in the experimental system could lead to observable differences in the transmission coefficients.

\subsection{\label{sub:coupling1}TLGR Coupling Coefficients $k_\mathrm{t}$ and $k_\mathrm{r}$}
\begin{figure*}
\begin{tabular}{cc}
(a)~\includegraphics[height=7.75cm]{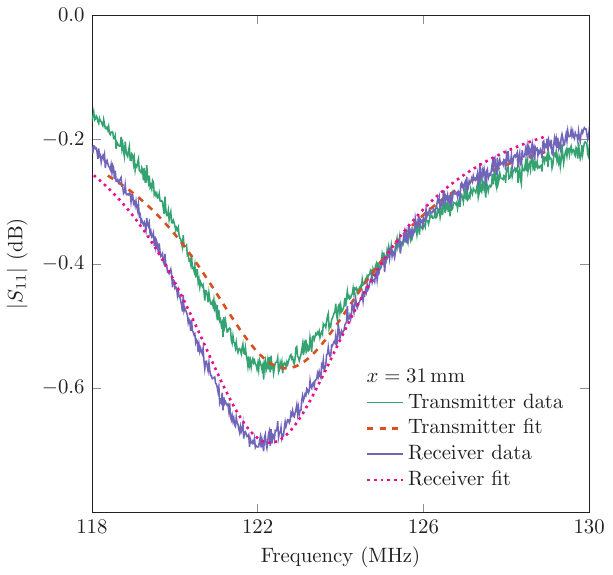} & (b)~\includegraphics[height=7.75cm]{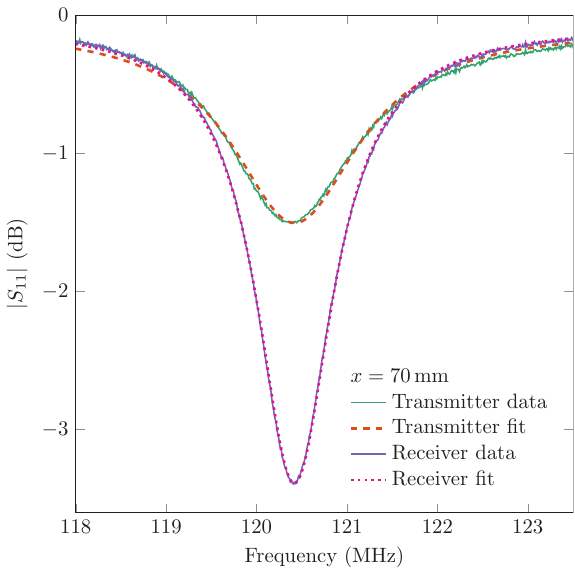}\\
~ & ~\\
(c)~\includegraphics[height=7.75cm]{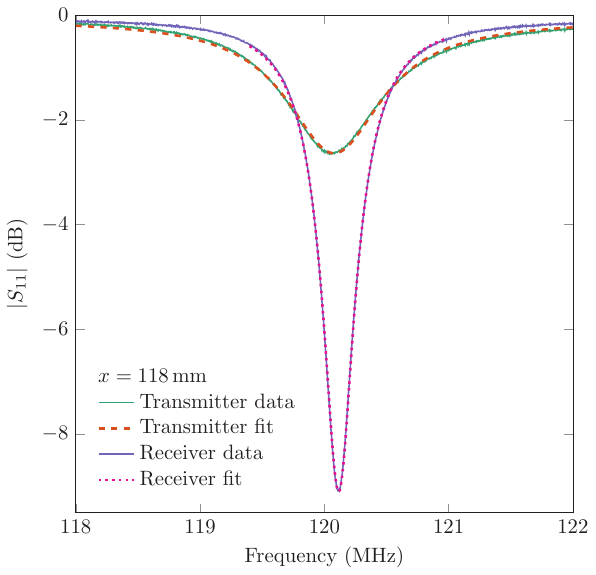} & (d)~\includegraphics[height=7.75cm]{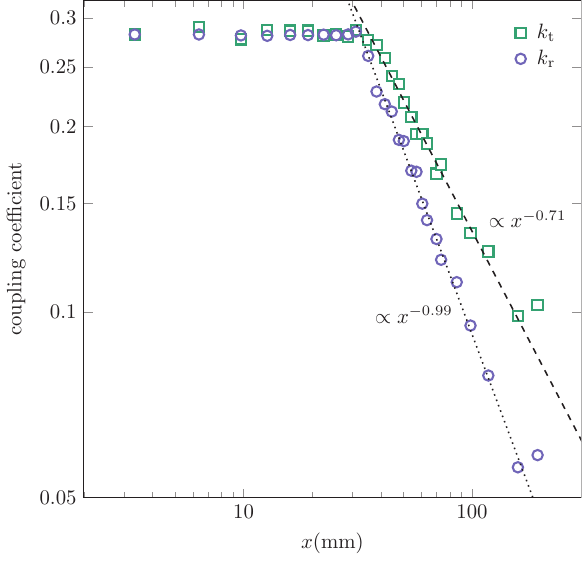}
\end{tabular}
\caption{\label{fig:S11}Fits to the frequency dependence of $\left\vert S_{11}\right\vert$ for isolated $\left(k\to 0\right)$ split-TLGRs.  The theoretical model for $\left\vert S_{11}\right\vert$ was generated using the equivalent circuit shown in Fig.~\ref{fig:circuits}(a) and the fits were used to extract values of the coupling coefficients $k_\mathrm{t}$ and $k_\mathrm{r}$. We show the fits for the transmitter and receiver when (a) $x=\SI{31}{\milli\meter}$, (b) $x=\SI{70}{\milli\meter}$, and (c) $x=\SI{118}{\milli\meter}$.  (d) The extracted values of $k_\mathrm{t}$ and $k_\mathrm{r}$ as a function of $x$.  For $x\le\SI{31}{\milli\meter}$ the coupling loops were oriented to achieve the maximum possible coupling and, therefore, $k_\mathrm{t}$ and $k_\mathrm{r}$ do not change over these separation distances.  Above $x=\SI{31}{\milli\meter}$, in order to maintain optimal power transfer efficiency, the coupling to the transmitting and receiving LGRs had to be decreased as $x$ was increased. The dashed and dotted lines are proportional to $x^{-0.71}$ and $x^{-0.99}$, respectively.}
\end{figure*}

For all of the $x$ values investigated using the split-TLGRs, after recording the $\left\vert S_{21}\right\vert$ frequency sweeps, the transmitter and receiver resonators were isolated from one another without changing the orientations of the coupling loops.  Next, the VNA was used to measure $\left\vert S_{11}\right\vert$ of the isolated resonators at both the transmitter and receiver coupling loop ports.  Example $\left\vert S_{11}\right\vert$ responses at three values of $x$ are shown in Figs.~\ref{fig:S11}(a) -- (c).  In Fig.~\ref{fig:S11}(a), the coupling loops were oriented to achieve maximum coupling such that the resonators, coupled to inductance $L_\alpha$ ($\alpha =\mathrm{s}$ or $\ell$), were overcoupled.  As a result, the values of $\left\vert Z\right\vert$ at the resonant frequency, determined from (\ref{eq:R}) and (\ref{eq:X}), were much greater than $Z_0=\SI{50}{\ohm}$ and the $\left\vert S_{11}\right\vert$ resonances were broad with minima that dipped only a fraction of a decibel below \SI{0}{\decibel}. In parts (b) and (c) of Fig.~\ref{fig:S11}, the coupling loops were rotated away from maximum coupling and the $\left\vert S_{11}\right\vert$ resonances sharpened and exhibited deeper minima.  For the split-TLGR IPT system, optimal power transfer was acheived with the transmitter resonator slightly more strongly coupled to $L_\alpha$ than the receiver resonator.  The $\left\vert S_{11}\right\vert$ data were fit to (\ref{eq:S11}) -- (\ref{eq:X}) based on the circuit model developed in Section~\ref{sec:circuit}.  

The loop inductances $L_\mathrm{s}$ and $L_\ell$ were determined from separate reflection coefficient measurements made after isolating the coupling loops from the TLGRs~\cite{Bobowski:2018}. The values obtained for the source and load loops were $L_\mathrm{s}=\SI[separate-uncertainty = true]{66.1\pm0.1}{\nano\henry}$ and $L_\ell=\SI[separate-uncertainty = true]{65.5\pm 0.1}{\nano\henry}$, respectively.  These measurements were compared to estimates made using an empirical expression for the inductance of a loop of wire
\begin{equation}
L_\alpha\approx \mu_0 r\left[\ln\left(\frac{8r}{a}\right)-2\right].\label{eq:L1}
\end{equation}
In this expression, $r$ is the loop radius and $a$ is the radius of the wire~\cite{Elliot:1993, Ramo:1993}.  Using $r=\SI{15}{\milli\meter}$ and $a=\SI{0.465}{\milli\meter}$ for our coupling loops made from UT-141 semi-rigid coaxial cable, (\ref{eq:L1}) gives an inductance of \SI{67.0}{\nano\henry} which is reasonably consistent with the experimental results.

The $\left\vert S_{11}\right\vert$ fits are all very good and the extracted parameters include the unloaded resonant frequencies $f_\beta$ and quality factors $Q_\beta$ of the transmitter and receiver LGRs as well as the coupling coefficients $k_\mathrm{t}$ and $k_\mathrm{r}$.  An offset parameter was also include in the fit to account for the fact that, away from resonance, the $\left\vert S_{11}\right\vert$ curves do not reach \SI{0}{\decibel}.  This effect is most clearly shown in Figs.~\ref{fig:S11}(a) and (b) and is likely due to losses along the lengths of the coaxial cables that lead up to the coupling loops.     

Table~\ref{tab:parameters} summarizes the parameters extracted from the $\left\vert S_{11}\right\vert$ fits shown in Fig.~\ref{fig:S11}.  For the transmitter, the $f_\mathrm{t}$ fit values were between \num{119.6} and \SI{120.1}{\mega\hertz} and $Q_\mathrm{t}$ varied from \num{850} to \num{1200}.  For the receiver we found a similar range of values for $f_\mathrm{r}$;  however, the quality factor was found to be slightly lower with $620 < Q < 850$.  Plots of the $x$-dependencies of $k_\mathrm{r}$ and $k_\mathrm{t}$ are shown in Fig.~\ref{fig:S11}(d).  For $x\lesssim \SI{31}{\milli\meter}$, the coupling loops were oriented parallel to the cross-section of the split-TLGR bore and $k_\mathrm{t}$ and $k_\mathrm{r}$ reached their maximum values.  As discussed in Section~\ref{sec:sims}, a substantial fraction of the total magnetic flux is contained within the resonator gap.  This observation explains why the maximum values of $k_\mathrm{t}$ and $k_\mathrm{r}$, being about \num{0.29}, are significantly less than one.  For $x>\SI{31}{\milli\meter}$, $k_\mathrm{t}$ and $k_\mathrm{r}$ decrease with increasing $x$.  For the transmitting side, we found $k_\mathrm{t}\propto x^{-0.71}$ and for the receiving side $k_\mathrm{r}\propto x^{-0.99}$.

\begin{table}
\caption{\label{tab:parameters} TGLR parameters associated with the  $\left\vert S_{21}\right\vert$ and $\left\vert S_{11}\right\vert$ fits shown in Figs.~\ref{fig:S21} and \ref{fig:S11}}
\centering
\begin{tabular}{l|cccc}
\hline\hline
\multicolumn{5}{c}{~}\\[-1.5ex]
\multicolumn{5}{c}{Isolated Transmitter $\left\vert S_{11}\right\vert$, Fig.~\ref{fig:S11}}\\[0.5ex]
\hline
\multicolumn{5}{c}{~}\\[-1.5ex]
\multicolumn{5}{c}{$L_\mathrm{s}=\SI{66.1}{\nano\henry}$,\qquad $L_\mathrm{t}=\SI{8.65}{\nano\henry}$}\\[0.5ex]
\hline
~ & ~\\[-1.5ex]
$x$ (\si{\milli\meter}) & \num{6.4} & \num{31} & \num{70} & \num{118}\\
$k_\mathrm{t}$ & \num{0.290} & \num{0.287} & \num{0.168} & \num{0.125}\\
$f_\mathrm{t}$ (\si{\mega\hertz}) & \num{119.73} & \num{120.09} & \num{119.58} & \num{119.607}\\
$Q_\mathrm{t}$ & \num{1020} & \num{1010} & \num{882} & \num{877}\\
offset & \num{0.018} & \num{0.017} & \num{0.010} & \num{0.009}\\[0.5ex]
\hline
\multicolumn{5}{c}{~}\\[-1.5ex]
\multicolumn{5}{c}{Isolated Receiver $\left\vert S_{11}\right\vert$, Fig.~\ref{fig:S11}}\\[0.5ex]
\hline
\multicolumn{5}{c}{~}\\[-1.5ex]
\multicolumn{5}{c}{$L_\ell=\SI{65.5}{\nano\henry}$,\qquad $L_\mathrm{r}=\SI{8.65}{\nano\henry}$}\\[0.5ex]
\hline
~ & ~\\[-1.5ex]
$x$ (\si{\milli\meter}) & \num{6.4} & \num{31} & \num{70} & \num{118}\\
$k_\mathrm{r}$ & \num{0.282} & \num{0.285} & \num{0.1314} & \num{0.0790}\\
$f_\mathrm{r}$ (\si{\mega\hertz}) & \num{119.60} & \num{119.79} & \num{119.90} & \num{119.930}\\
$Q_\mathrm{r}$ & \num{850} & \num{740} & \num{618} & \num{680}\\
offset & \num{0.015} & \num{0.012} & \num{0.008} & \num{0.008}\\[0.5ex]
\hline
\multicolumn{5}{c}{~}\\[-1.5ex]
\multicolumn{5}{c}{Coupled Transmitter/Receiver IPT $\left\vert S_{21}\right\vert$, Fig.~\ref{fig:S21}}\\[0.5ex]
\hline
~ & ~\\[-1.5ex]
$x$ (\si{\milli\meter}) & \num{6.4} & \num{31} & \num{70} & \num{118}\\
$k$ & \num{0.192} & \num{0.0383} & \num{0.0093} & \num{0.0026}\\
$b_\mathrm{t}$ & \num{1.083} & \num{1.004} & \num{0.996} & \num{0.999}\\
$b_\mathrm{r}$ & \num{0.989} & \num{0.983} & \num{1.002} & \num{1.000}\\[0.5ex]
\hline\hline
\end{tabular}
\end{table}

\subsection{\label{sub:coupling2}TLGR Coupling Coefficients $k$}
With all of $L_\mathrm{s}$, $L_\ell$, $f_\mathrm{t}$, $f_\mathrm{r}$, $Q_\mathrm{t}$, $Q_\mathrm{r}$, $k_\mathrm{t}$, and $k_\mathrm{r}$ known, the system of equations (\ref{eq:sys1}) -- (\ref{eq:sys4}) can be solved for $I_\mathrm{r}$ and the only unknown parameters in \mbox{$\left\vert S_{21}\right\vert =2\left\vert I_\ell/V_\mathrm{s}\right\vert Z_0$} are the TLGR inductances, $L_\mathrm{t}$ and $L_\mathrm{r}$, and the coupling coefficient $k$. The effective inductance of a complete TLGR was calculated in \cite{Bobowski:2016}.  Assuming that the magnetic fields outside the bore are negligible, the split-TLGR has twice the inductance of a full TLGR such that
\begin{equation}
L_\beta\approx 2\mu_0 r_1\left[1-\sqrt{1-\left(\dfrac{r_0}{r_1}\right)^2}\right].\label{eq:Lbeta}
\end{equation} 
Using the dimensions in Fig.~\ref{fig:LGRs}(c), (\ref{eq:Lbeta}) gives \mbox{$L=\SI{8.65}{\nano\henry}$} which was the value assumed for both $L_\mathrm{t}$ and $L_\mathrm{r}$. We found that the quality of the fits were very sensitive to the values of $f_\mathrm{t}$ and $f_\mathrm{r}$.  To ensure high-quality fits, we introduced additional fitting parameters $b_\mathrm{t}$ and $b_\mathrm{r}$ to scale the values of $f_\mathrm{t}$ and $f_\mathrm{r}$, respectively.  On average, the scaling factors changed $f_\mathrm{t}$ and $f_\mathrm{r}$ by only \SI{1.6}{\percent} and \SI{0.8}{\percent}, respectively.  

The fits to $\left\vert S_{21}\right\vert$ measurements are generally good and  examples are shown in Fig.~\ref{fig:S21}(a) -- (c).  The fit function underestimates the measured data below $f_0$ and overestimates it above $f_0$.  The mismatch between the theoretical model and experimental data becomes more exaggerated as $x$ is decreased.  These same features were observed in the analysis by Sample {\it et al.}\ in \cite{Sample:2011}.  Those authors improved the quality of their fits by introducing two more fit parameters to account for the cross-couplings between the source loop and receiver and between the load loop and transmitter.  We have not included cross-couplings in our analysis.  The additional fit parameters will undoubtedly improve the quality of the fit, however, it is not clear that it results in more reliable values for the coupling coefficient $k$.  Furthermore, one would always expect the circuit models to have some deficiencies due to the fact that the transmitter and receiver resonators, whatever their geometry, are not lumped-element $LRC$ circuits.       

The $x$-dependence of $k$, determined from the $\left\vert S_{21}\right\vert$ fits, is shown in Fig.~\ref{fig:S21}(d).  The data are plotted as a function of $x^{-1/2}$ to highlight that, in the strong coupling limit (small $x$), $k\propto x^{-1/2}$.  In the overcoupled regime, where $\left\vert S_{21}\right\vert$ splits into a double resonance, the coupling coefficient $k$ can also be estimated using
\begin{equation}
k=\frac{f_2^2-f_1^2}{f_2^2+f_1^2},\label{eq:k}
\end{equation}
where $f_1$ and $f_2$ are the resonant frequencies of the even and odd modes, respectively~\cite{Hong:2001}.  For the $x=\SI{6.4}{\milli\meter}$ data shown in Fig.~\ref{fig:S21}(a), (\ref{eq:k}) gives $k\approx\num{0.22}$ which is in reasonably good agreement with \num{0.192} obtained from the fit.

\begin{figure}
\includegraphics[height=8cm]{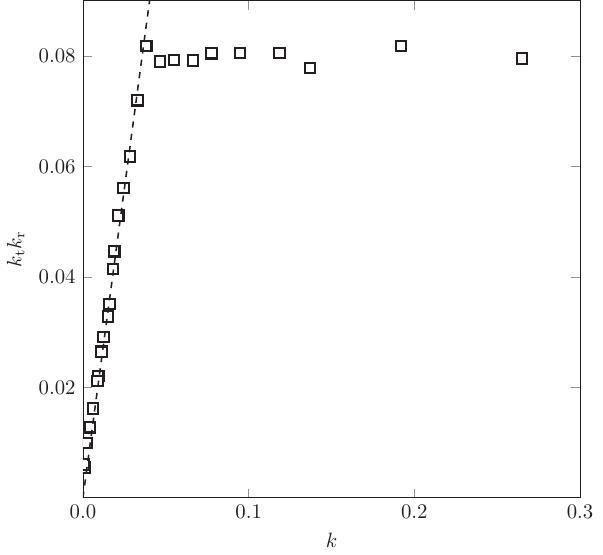}
\caption{\label{fig:ktkrvsk}The product $k_\mathrm{t}k_\mathrm{r}$ as a function of $k$.  For weak coupling (\mbox{$k\lesssim 0.04$}), optimal power transfer efficiency corresponds to the condition that the $k_\mathrm{t}k_\mathrm{r}$ product  is proportional to $k$.  The dashed line is \mbox{$k_\mathrm{t}k_\mathrm{r}=2.24k$}.  When \mbox{$k\gtrsim 0.04$} (corresponding to $x<\SI{31}{\milli\meter}$), the $k_\mathrm{t}k_\mathrm{r}$ value saturates as both the source and load coupling loops are already oriented to achieve maximum coupling.}
\end{figure}

Many authors have shown that four-coil IPT systems achieve optimal power transfer efficiency when the coupling coefficients are tuned such that $k\propto k_\mathrm{t}k_\mathrm{r}$~\cite{Sample:2011, Cheon:2011, Duong:2011, Duong:2015}.  Figure~\ref{fig:ktkrvsk} shows a plot of $k_\mathrm{t}k_\mathrm{r}$ as a function of $k$ for the split-TLGR system based on the $\left\vert S_{11}\right\vert$ and $\left\vert S_{21}\right\vert$ analyses described in Sections~\ref{sub:coupling1} and \ref{sub:coupling2}, respectively.  For weak coupling, before $k_\mathrm{t}$ and $k_\mathrm{r}$ have reached their maximum values, the linear relation between the $k_\mathrm{t}k_\mathrm{r}$ product and the coupling coefficient $k$ is clearly observed. 

With all of the TLGR IPT system parameters known (Table~\ref{tab:parameters}), the equivalent circuit model of Fig.~\ref{fig:circuits}(b) can be used to calculate the currents in transmit and receive resonators.  At a frequency of \SI{121}{\mega\hertz}, we found \mbox{$\left\vert I_\mathrm{t}\right\vert=\left\vert I_\mathrm{r}\right\vert =\SI{262}{\milli\ampere}$}.  From Amp\`ere's law, the magnetic field strength at the inner radius of the split-TLGR bore is given by \mbox{$B=\mu_0 \left\vert I_\beta\right\vert/\left[\pi\left(r_1-r_0\right)\right]=\SI{3.7}{\micro\tesla}$} which is in reasonably good agreement with the simulation result of \SI{3.2}{\micro\tesla} shown in Fig.~\ref{fig:COMSOLscan}(b).  An estimate of the electric field strength can be made by calculating the voltage across the gap and then dividing by the gap height: $E=\left\vert I_\beta\right\vert/\left(\omega C_\beta t\right)=I_\beta\omega_\beta^2 L_\beta/\left(\omega t\right)$, where $\omega_\beta^2=1/\left(L_\beta C_\beta\right)$.  At \SI{121}{\mega\hertz}, the result is $E=\SI{17}{\kilo\volt/m}$ which is again in reasonable agreement with the simulation result of \SI{15}{\kilo\volt/m}.   

\subsection{\label{sub:bandwidth}CLGR and TLGR IPT System Bandwidths}
\begin{figure*}
\begin{tabular}{cc}
(a)~\includegraphics[height=7.75cm]{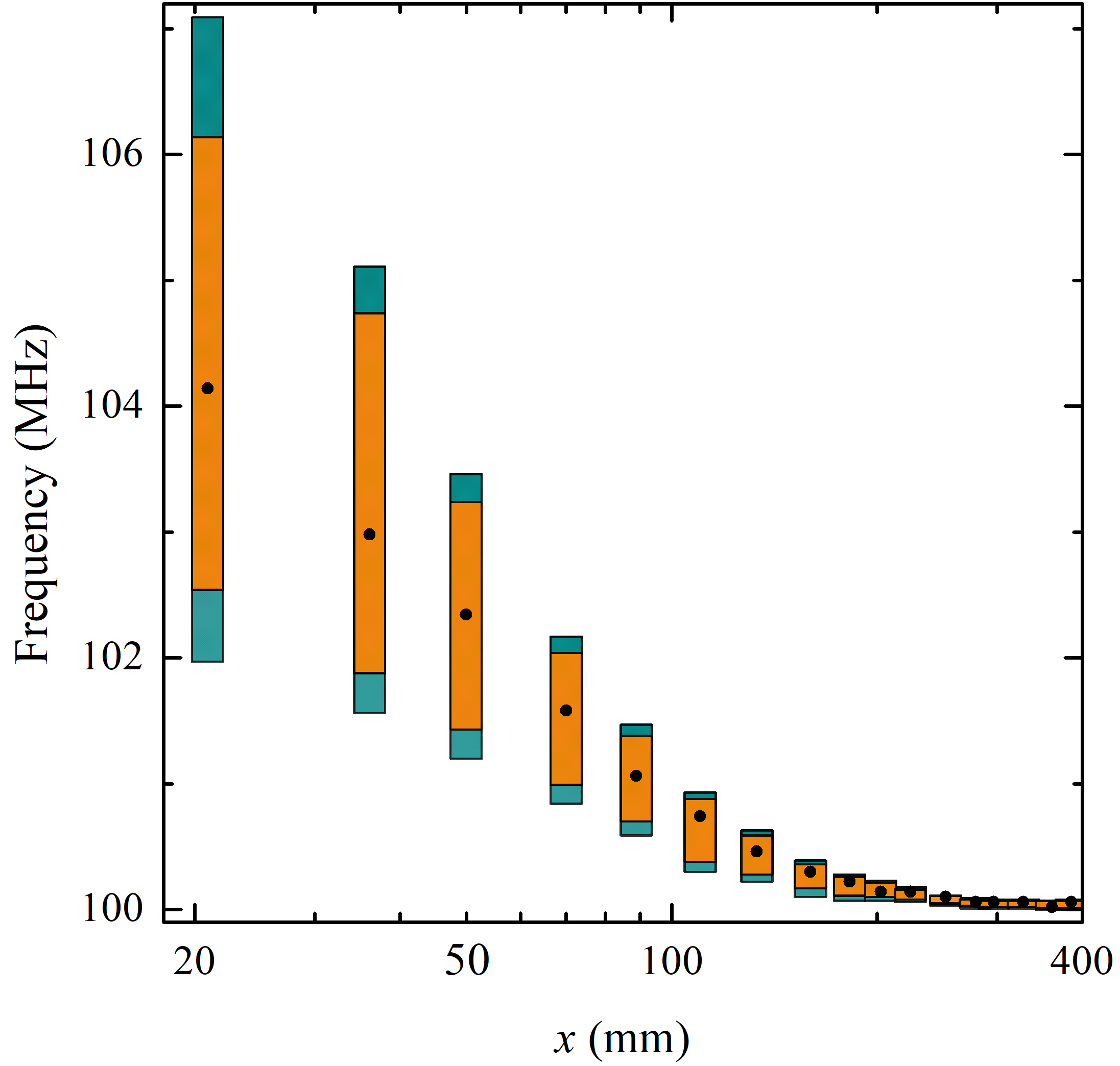} & (b)~\includegraphics[height=7.75cm]{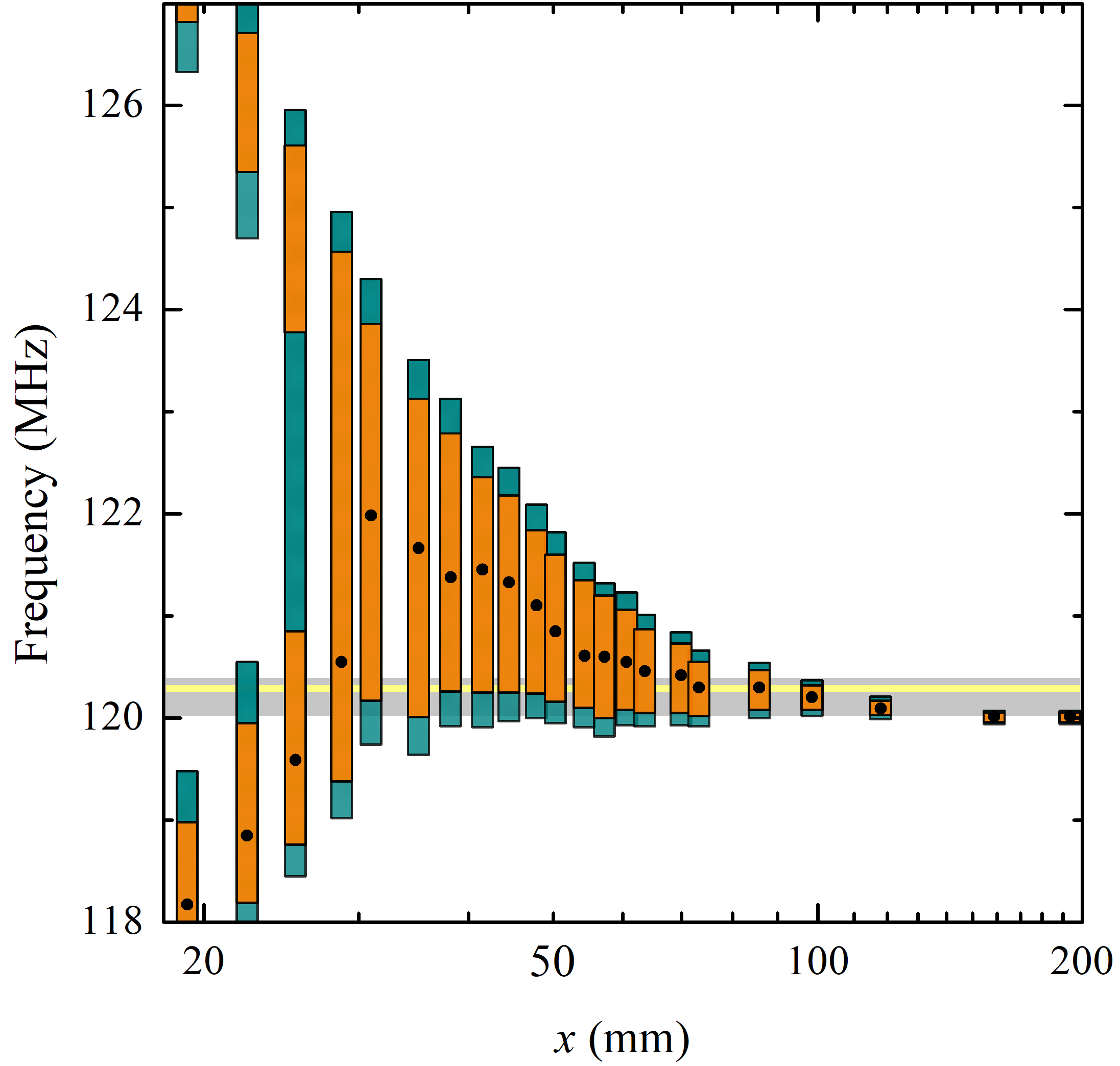}
\end{tabular}
\caption{\label{fig:bandwidth}Bandwidth analysis of the LGR IPT systems versus separation distance $x$.  The dots show the frequency of peak $\left\vert S_{21}\right\vert$ as a function of $x$.  The orange (green) bars represent the range of frequencies for which $\left\vert S_{21}\right\vert$ is within \SI{0.22}{\decibel} (\SI{0.46}{\decibel}) of its peak value. (a) For the CLGR system, the top and bottom ends of the \num{0.22} and \SI{0.46}{\decibel} frequency ranges both change with $x$. (b) For the TLGR system the bottom ends of the frequency ranges are approximately constant when $x>\SI{31}{\milli\meter}$.  As a result, there exists a band of frequencies for which $\left\vert S_{21}\right\vert$ remains close to the peak value as $x$ is varied.  The horizontal grey bar, \SI{0.4}{\mega\hertz} wide, represents the band of frequencies for which $\left\vert S_{21}\right\vert$ is always within \SI{0.46}{\decibel} of its peak value when $22\le x\le \SI{100}{\milli\meter}$.  The yellow bar, \SI{0.1}{\mega\hertz} wide, shows the bandwidth for the \SI{0.22}{\decibel} case when $25\le x\le \SI{100}{\milli\meter}$.}
\end{figure*}

To produce the plots of peak $\left\vert S_{21}\right\vert$ versus $x$ shown in Fig.~\ref{fig:peakS21} for the CLGR and split-TLGR systems, the coupling loop positions/orientations were tuned to optimize the power transfer efficiency at each value of $x$.  A drawback of this approach is that the frequency of peak power transfer efficiency $f_\mathrm{p}$ varies with $x$.  The data points in Figs.~\ref{fig:bandwidth}(a) and (b) show the $x$-dependencies of the peak frequency for the CLGR and split-TLGR systems.  Excluding the overcoupled region of the TLGR data, $f_\mathrm{p}$ was found to be a nearly linear function of $x^{-1}$ for both systems when $x<x_0$.  Because low-frequency ISM bands are typically only tens of kilohertz wide, practical IPT systems are restricted to narrow bandwidths limiting the application of frequency agile sources. Therefore, the performance under fixed-frequency source conditions was evaluated.

The green vertical bars in Fig.~\ref{fig:bandwidth} represent the range of frequencies for which $\left\vert S_{21}\right\vert$ is within \SI{0.458}{\decibel} of its peak value, or the power transfer efficiency is no worse than \SI{90}{\percent} of the optimal efficiency.  For example, the $x=\SI{31}{\milli\meter}$ data shown in Fig.~\ref{fig:S21}(b) have a maximum $\left\vert S_{21}\right\vert$ value of \SI{-0.306}{\decibel} at \SI{122.0}{\mega\hertz}.  $\left\vert S_{21}\right\vert$ drops to \SI{-0.764}{\decibel} at \num{119.7} and \SI{124.3}{\mega\hertz} which defines the bandwidth spanned by the vertical green bar in Fig.~\ref{fig:bandwidth}(b).  This bandwidth shrinks as $x$ increases.  For example, the $x=\SI{70}{\milli\meter}$ data shown in Fig.~\ref{fig:S21}(c) have a bandwidth of \SI{0.91}{\mega\hertz} which is only \SI{20}{\percent} of the bandwidth at $x=\SI{31}{\milli\meter}$.  Once $x>x_0$, the bandwidth rapidly shrinks for both the CLGR and split-TLGR systems.  The orange bars show the bandwidth at different $x$ when $\left\vert S_{21}\right\vert$ is required to be within \SI{0.223}{\decibel} of its peak value, which corresponds to a power transfer efficiency that is no worse than \SI{95}{\percent} of peak efficiency. 

The most striking difference between Figs.~\ref{fig:bandwidth}(a) and (b) is the $x$-dependence of the lower-limit of the bandwidth.  For the CLGR, both the upper and lower limits of the bandwidth decrease approximately linearly with $x^{-1}$.  As a result, there is no fixed frequency that will maintain high power-transfer efficiency for a wide range of $x$-values.  For the split-TLGR system, while the upper limit of the bandwidth does vary linearly with $x^{-1}$, the lower limit remains nearly constant for $x>\SI{31}{\milli\meter}$.  This property of the split-TLGR system was unexpected and shows that efficient power transfer can be obtained at a fixed frequency over a wide range of $x$ values.  The horizontal region shaded gray in Fig.~\ref{fig:bandwidth}(b) highlights the range of frequencies, \SI{0.4}{\mega\hertz} wide, that is always within \SI{0.458}{\decibel} of the peak $\left\vert S_{21}\right\vert$ when $\num{22}\le x\le\SI{100}{\milli\meter}$.  The yellow bar defines the band of frequencies, \SI{0.1}{\mega\hertz} wide, for which $\left\vert S_{21}\right\vert$ is always within \SI{0.223}{\decibel} of the peak value when $\num{25}\le x\le\SI{100}{\milli\meter}$.  As discussed in Section~\ref{sub:expt1}, the lower limit of the $x$ ranges could be extended by increasing $L_\mathrm{s}$ and $L_\ell$ so as to avoid the frequency splitting associated with overcoupled IPT systems. 

\subsection{\label{sub:HP} Operation of the TLGR IPT System at \SI{32}{\watt}}
All of the measurements described up to this point were taken using a VNA that supplies \SI{-17}{dBm} (\SI{20}{\micro\watt}) of source power.  We now present a characterization of the split-TLGR IPT system at \SI{45}{dBm} (\SI{32}{\watt}).  The source signal was supplied by a Rhode \& Schwarz SMY02 signal generator followed by a Mini-Circuits LZY-22+ power amplifier.  To measure the forward power from the amplifier, a OSR Broadcast Research C21A8 directional coupler, with a coupling of \SI{40}{\decibel}, was inserted between the amplifier and the source coupling loop.  The load loop was terminated by a \SI{50}{\ohm} load with a \SI{100}{\watt} power rating.  A second directional coupler was used to measure the power delivered to the load.  The incident power $P_0$ and transmitted power $P_\mathrm{L}$ were monitored using a Keysight E4417A dual-channel power meter paired with Keysight E9320 power sensors.  A simple LabVIEW program was written to step through a range of frequencies and record $P_0$, $P_\mathrm{L}$, and \mbox{$\left\vert S_{21}\right\vert=10\log\left(P_\mathrm{L}/P_0\right)$} in decibels.  

Measurements were taken at six different values of $x$ between \num{6} and \SI{80}{\milli\meter}.  For each separation distance, the coupling loop orientations were first optimized using the VNA before completing frequency sweeps at higher power.  For all six values of $x$, the VNA and high-power $\left\vert S_{21}\right\vert$ measurements were in good agreement.  The small discrepancies that were observed are attributed to directional coupler insertion loss and imperfect power sensor calibrations.  Figure~\ref{fig:S21}(b) shows an overlay of the high-power and VNA $\left\vert S_{21}\right\vert$ measurements for the critically-coupled case when $x=\SI{31}{\milli\meter}$.  The most significant difference between the two data sets is the peak value of $\left\vert S_{21}\right\vert$.  The high-power data reach a maximum of \SI{-0.598}{\decibel} compared to \SI{-0.306}{\decibel} for the VNA data.

\section{\label{sec:discussion}Discussion}
The IPT system using CLGRs as transmitter and receiver resulted in a large spatial bandwidth $x_0$ and a competitive figure of merit $x_0/s$.  Therefore, this system is well suited to applications which require large separations between the transmitter and receiver.  The CLGR system strongly confines electric fields within the gap of the resonator thereby desensitizing the power transfer efficiency to the presence of extraneous dielectric objects.  The magnetic fields, on the other hand, occupy the entire volume of space surrounding the CLGRs which enables some degree of omnidirectional IPT.

The system characterized in this work operated at a \SI{100}{\mega\hertz}.  Reducing the resonant frequency by either increasing $d_\mathrm{c}$ or inserting a high-$\varepsilon_\mathrm{r}$ and low-loss dielectric in the gap, would increase the ratio $\lambda/d_\mathrm{c}$ and diminish radiative losses such that an enhancement of the figure of merit would be expected.  

It is worth pointing out that changing the length of the CLGRs has only a small effect on the resonant frequency~\cite{Madsen:2020}.  Therefore, it is possible to select the CLGR transmitter length that produces the desired magnetic field profile outside the resonator.  For example, a short CLGR could be used to produce magnetic fields patterns similar to those generated by helical and spiral resonators.

Although the figure of merit of the split-TLGR IPT system was low compared to that of the CLGR system, it has other practical advantages that make it an attractive option for some applications.  For example, the system can be operated efficiently at a fixed frequency for a wide range of transmitter-receiver distances and the unique geometry offers a way to manipulate the spatial distribution of the magnetic fields.  In particular, the magnetic field strength everywhere outside the resonators, excluding the region of space directly between the transmitter and receiver, is very low.  These features have the potential to make the system a good choice for some specialized applications in which the lateral alignment of the transmitter and receiver can be maintained.

For example, because the electric and magnetic fields of the split-TLGR system are restricted to only a few well-defined regions of space, various insulating and conducting objects can be situated nearby the resonators without adversely affecting the power transfer efficiency.  Lossy dielectrics can be placed anywhere except within the narrow gap of resonator.  Conducting objects can likewise be located anywhere except within the volumes of space that connect the bores of the transmit and receive resonators.  A possible application is IPT through a slab of concrete embedded with reinforcing steel (rebar).  As long as the rebar does not pass through the volumes of space connecting the transmitter and receiver bores, efficient energy exchange is possible.  Other IPT designs, with magnetic fields that occupy the entire volume of space surrounding the resonators, would suffer losses due to eddy currents induced in the steel bars.  A similar application would be efficient IPT through a wall that has plumbing, duct work, and/or electrical conduits running parallel to its surfaces.

As discussed in Section~\ref{sec:LGRs}, the electric field confinement provided by LGRs make them good candidates for IPT applications in saltwater.  Even with the electric fields isolated from the conducting medium, a magnetic power dissipation proportional to $\sigma\omega_0^2 B^2$ will limit the power transfer efficiency~\cite{Wandinger:2021}.  If the conducting medium is excluded from LGR bores using seals or another means, the magnetic power dissipation will be due only to the fields outside the resonators.  In the case of the split-TLGRs, the external fields are only substantial between the transmitter and receiver such that losses associated with the conducting medium outside of this region would be negligible.  Low-frequency split-TLGRs could be used, for example, to wirelessly recharge autonomous underwater vehicles (AUVs) at a charging base beneath the ocean or sea surface.      

\section{\label{sec:conclusions}Conclusions}
We have implemented efficient IPT systems that operate near \SI{100}{\mega\hertz} using inductively-coupled LGR transmitters and receivers.  The LGRs are high-$Q$ structures that are small in size compared to the free-space wavelength at resonance.  With a constraint on the maximum size of the resonators, our designs minimize the resonance frequency.  Impedance matching is achieved by relatively simple adjustments to the position/orientation of source and load coupling loops.  Lumped-element equivalent circuits were used to develop theoretical models that could be fit to measured scattering parameters and used to extract the three relevant coupling coefficients.

We designed and investigated IPT systems using both CLGRs and split-TLGRs.  Although the CLGR system allowed for efficient power transfer at greater separation distances between transmitter and receiver, the split-TLGR system offered two significant advantages: (1) there exists an operating bandwidth for near-optimal power transfer that spans a wide range of transmitter-receiver separation distances and (2) there is minimal external magnetic flux outside of the region directly between the transmitter and receiver making the system insensitive to the surrounding environment.

With the source and load coupling loop orientations tuned to achieve maximum power transfer, the transmitter and receiver TLGRs were isolated from one another and fits to the frequency dependence of $\left\vert S_{11}\right\vert$ were used to determine $k_\mathrm{t}$ and $k_\mathrm{r}$.  Even with the loop orientations set to achieve maximum coupling, the largest coupling coefficients obtained were only 0.29.  COMSOL simulations showed the presence of substantial magnetic flux within the gap of the TLGRs which explains why coupling loops suspended within the bore of the resonator are not able to achieve stronger couplings.

Using the experimentally-determined values of $k_\mathrm{t}$ and $k_\mathrm{r}$, fits to the frequency dependence of $\left\vert S_{21}\right\vert$ were used to extract the coupling coefficient $k$ between transmitter and receiver.  At strong coupling (small $x$), $k$ followed a $x^{-1/2}$ dependence.  Furthermore, as predicted by equivalent circuit models, we found that optimal power transfer is achieved when the condition $k_\mathrm{t}k_\mathrm{r}\propto k$ is satisfied.

In future work, we plan to operate the LGR-based IPT designs at lower frequencies.  One approach to lower the frequency is to insert high-$\varepsilon_\mathrm{r}$ and low-loss dielectrics within the gap of the resonators.  We will also investigate IPT through a conducting medium, such as saltwater, using LGR transmitters and receivers.

\section*{Acknowledgment}
We thank NSERC for providing funding for this project and we are grateful for the fabrication of the LGRs by the UBC Okanagan machine shop.

\ifCLASSOPTIONcaptionsoff
  \newpage
\fi



\bibliographystyle{IEEEtran}
%\bibliography{TSRR_refs}


\vfill
\newpage

\begin{IEEEbiography}[{\includegraphics[width=1in,height=1.25in,clip,keepaspectratio]{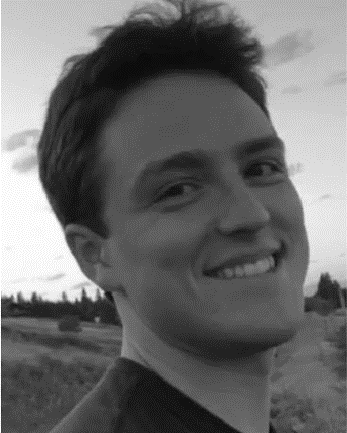}}]{David M. Roberts}
  was born in Cranbrook, Canada in April 1997. He received a B.Sc. degree (Hons.) in physics from the University of British Columbia, Kelowna, BC in 2020. He is now pursuing an M.Sc. in physics at the University of British Columbia, Vancouver, BC. His research interests include using resonant cavities in silicon to couple spin-qubits. 
\end{IEEEbiography}

\begin{IEEEbiography}[{\includegraphics[width=1in,height=1.25in,clip,keepaspectratio]{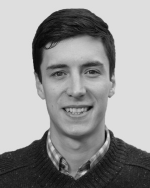}}]{Aaron P. Clements}
  was born in Nelson, New Zealand in 1993. He received the B.Sc. degree with honors in physics from the University of British Columbia, Kelowna, BC in 2016.
As a research assistant at UBC's Okanagan campus, he has worked on a range of research topics which include harnessing non-Newtonian fluid rheology in a novel pump design and non-contact detection of high-voltage hazards. For his honors project he used a toroidal split-ring resonator to implement an electron spin resonance experiment at \SI{1}{GHz}. He is currently employed as a research assistant in the School of Engineering at UBC's Okanagan campus, focused on resonant structure design for wireless power applications.
\end{IEEEbiography}

\begin{IEEEbiography}[{\includegraphics[width=1in,height=1.25in,clip,keepaspectratio]{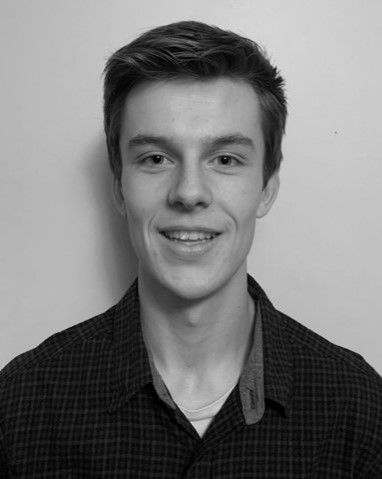}}]{Rowan McDonald}
 was born in Salmon Arm, Canada, in 1999. He is currently a B.A.Sc. student at the University of British Columbia, Vancouver, BC, Canada.

Since 2019, he has been a research assistant at the University of British Columbia. He has been involved in multiple projects including the design of a capacitive wireless power transfer system and the investigation of transistor aging and self-healing in nanoscale devices.
\end{IEEEbiography}

\vfill
\newpage

\begin{IEEEbiography}[{\includegraphics[width=1in,height=1.25in,clip,keepaspectratio]{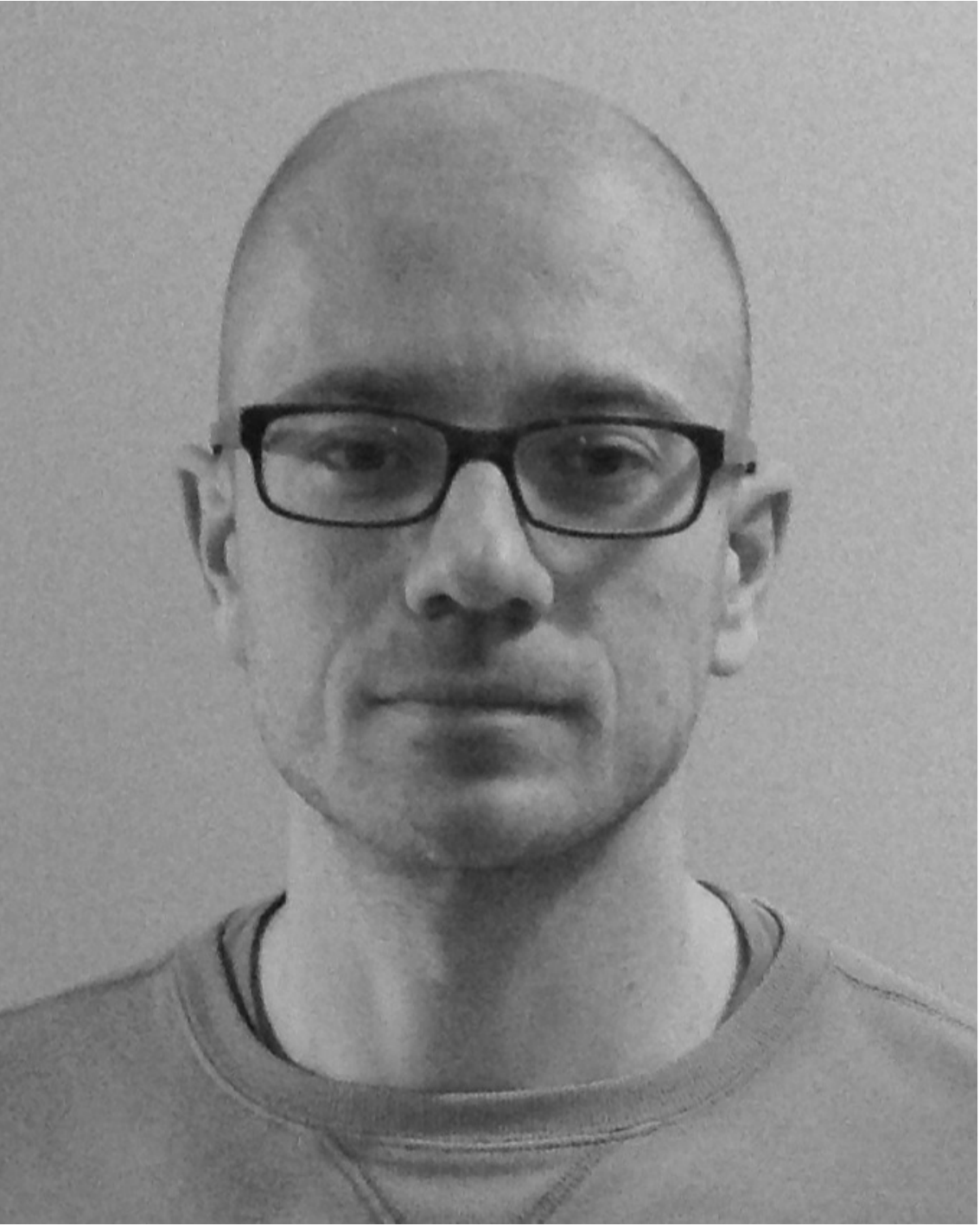}}]{Jake S. Bobowski}
was born in Winnipeg, Canada on March 9, 1979.  He received a B.Sc. degree in physics from the University of Manitoba, Canada in 2001.  He was awarded M.Sc. and Ph.D. degrees in physics from the University of British Columbia, Canada in 2004 and 2010, respectively.  From 2011 to 2012, he was a postdoctoral fellow in the RF and Microwave Technology Research Laboratory in the Department of Electrical Engineering, and is now an Associate Professor of Teaching in physics, at the Okanagan campus of the University of British Columbia, Canada.  He is interested in resonator design and developing custom microwave techniques to characterize the electromagnetic properties of a wide range of materials.
\end{IEEEbiography}

\begin{IEEEbiography}[{\includegraphics[width=1in,height=1.25in,clip,keepaspectratio]{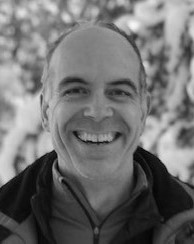}}]{Thomas~Johnson}
(M'02) received the B.A.Sc. degree in electrical engineering from The University of British Columbia in 1987. In 2001 and 2007, he received  M.A.Sc. and Ph.D. degrees from Simon Fraser University where he focused on research in RF and microwave power amplifiers. He is currently an Associate Professor in the School of Engineering at UBC. He leads the RF and Microwave Technology Research Laboratory at UBC, solving applied problems in the area of radio frequency and microwave circuits and systems. His research interests include the design of radio frequency circuits and systems,  wireless power systems, wireless sensors and industrial applications of RF/microwave power. Before joining UBC in 2009, he was a technical lead in a number of high-tech companies including PulseWave RF, ADC Telecommunications, and Norsat International.
\end{IEEEbiography}
\vfill

%\newpage


%\vfill


\end{document}